\documentclass[twocolumn]{aastex61}

\usepackage{epstopdf}
\usepackage{multirow}
\usepackage{threeparttable}
\usepackage{natbib}
\usepackage{float}
\usepackage{amsmath}
\newcommand{\unit}[1]{\ensuremath{\, \mathrm{#1}}}

\newcommand{\Msol}{\hbox{$\mathrm{M_\odot}$}}
\newcommand{\msol}{\hbox{$\mathrm{M_\odot}$}}

\newcommand{\redshift}{z}

\newcommand{\editone}[1]{{#1}}

\begin{document}
\title{Effect of local environment and stellar mass on galaxy quenching and morphology at $0.5<\protect \redshift<2.0$
\footnote{T\lowercase{his paper includes data gathered with the 6.5 meter} M\lowercase{agellan} T\lowercase{elescopes located at} L\lowercase{as} C\lowercase{ampanas} O\lowercase{bservatory}, C\lowercase{hile}.}}

\correspondingauthor{Lalitwadee Kawinwanichakij}
\email{kawinwanichakij@physics.tamu.edu}
\author[0000-0003-4032-2445]{Lalitwadee Kawinwanichakij}
\affiliation{Department of Physics and Astronomy, Texas A\&M University, College
Station, TX, 77843-4242 USA}
\affiliation{George P.\ and Cynthia Woods Mitchell Institute for
  Fundamental Physics and Astronomy, Texas A\&M University, College
  Station, TX, 77843-4242}
\affiliation{LSSTC Data Science Fellow}  
\author[0000-0001-7503-8482]{Casey Papovich}
\affiliation{Department of Physics and Astronomy, Texas A\&M University, College
Station, TX, 77843-4242 USA}
\affiliation{George P.\ and Cynthia Woods Mitchell Institute for
  Fundamental Physics and Astronomy, Texas A\&M University, College
  Station, TX, 77843-4242}
  
\author[0000-0003-0341-8827]{Ryan F. Quadri}
\affiliation{Department of Physics and Astronomy, Texas A\&M University, College
Station, TX, 77843-4242 USA}
\affiliation{George P.\ and Cynthia Woods Mitchell Institute for
  Fundamental Physics and Astronomy, Texas A\&M University, College
  Station, TX, 77843-4242}
\affiliation{Mitchell Astronomy Fellow}

\author[0000-0002-3254-9044]{Karl Glazebrook}
\affiliation{Centre for Astrophysics and Supercomputing, Swinburne University, Hawthorn, VIC 3122, Australia}

\author{Glenn G. Kacprzak}
\affiliation{Centre for Astrophysics and Supercomputing, Swinburne University, Hawthorn, VIC 3122, Australia} 

\author{Rebecca J. Allen}
\affiliation{Centre for Astrophysics and Supercomputing, Swinburne University, Hawthorn, VIC 3122, Australia}

\author[0000-0002-5564-9873]{Eric F. Bell}
\affiliation{Department of Astronomy, University of Michigan, 1085 South University Ave., Ann Arbor, MI 48109-1107}

\author[0000-0002-5009-512X]{Darren J. Croton}
\affiliation{Centre for Astrophysics and Supercomputing, Swinburne University, Hawthorn, VIC 3122, Australia}

\author[0000-0003-4174-0374]{Avishai Dekel}
\affiliation{Racah Institute of Physics, The Hebrew University, Jerusalem 91904 Israel}

\author[0000-0001-7113-2738]{Henry C. Ferguson}
\affiliation{Space Telescope Science Institute, 3700 San Martin Dr., Baltimore, MD 21218, USA}

\author[0000-0001-6003-0541]{Ben Forrest}
\affiliation{Department of Physics and Astronomy, Texas A\&M University, College
Station, TX, 77843-4242 USA}
\affiliation{George P.\ and Cynthia Woods Mitchell Institute for
  Fundamental Physics and Astronomy, Texas A\&M University, College
  Station, TX, 77843-4242}

\author[0000-0001-9440-8872]{Norman A. Grogin}
\affiliation{Space Telescope Science Institute, 3700 San Martin Dr., Baltimore, MD 21218, USA}

\author{Yicheng Guo}
\affiliation{UCO/Lick Observatory, Department of Astronomy and Astrophysics,
University of California, Santa Cruz, CA, USA}

\author[0000-0002-8360-3880]{Dale D. Kocevski}
\affiliation{Department of Physics and Astronomy, Colby College, Waterville,
Maine 04901, USA}

\author[0000-0002-6610-2048]{Anton M. Koekemoer}
\affiliation{Space Telescope Science Institute, 3700 San Martin Dr., Baltimore, MD 21218, USA}

\author[0000-0002-2057-5376]{Ivo Labb\'{e}}
\affiliation{Leiden Observatory, Leiden University, P.O. Box 9513, 2300 RA Leiden, The Netherlands}

\author[0000-0003-1581-7825]{Ray A. Lucas}
\affiliation{Space Telescope Science Institute, 3700 San Martin Dr., Baltimore, MD 21218, USA}

\author[0000-0003-2804-0648]{Themiya Nanayakkara}
\affiliation{Centre for Astrophysics and Supercomputing, Swinburne University, Hawthorn, VIC 3122, Australia}
\affiliation{Leiden Observatory, Leiden University, P.O. Box 9513, 2300 RA Leiden, The Netherlands}

\author[0000-0001-5185-9876]{Lee R. Spitler}  
\affiliation{Australian Astronomical Observatory, P.O. Box 915, North Ryde, NSW 1670, Australia} 
\affiliation{Research Centre for Astronomy, Astrophysics \& Astrophotonics, Macquarie University, Sydney, NSW 2109, Australia}
\affiliation{Department of Physics \& Astronomy, Macquarie University, Sydney, NSW 2109, Australia}

\author[0000-0001-5937-4590]{Caroline M. S. Straatman}
\affiliation{Max-Planck Institut f\"ur Astronomie, K\"onigstuhl 17, D$-$69117, Heidelberg, Germany}

\author[0000-0001-9208-2143]{Kim-Vy H. Tran} 
\affiliation{Department of Physics and Astronomy, Texas A\&M University, College
Station, TX, 77843-4242 USA}
\affiliation{George P.\ and Cynthia Woods Mitchell Institute for
  Fundamental Physics and Astronomy, Texas A\&M University, College
  Station, TX, 77843-4242}
\author[0000-0003-2008-1752]{Adam Tomczak}  
\affiliation{Department of Physics, UC Davis, Davis, CA 95616, USA}  

\author[0000-0002-8282-9888]{Pieter van Dokkum}
\affiliation{Astronomy Department, Yale University, New Haven, CT 06511, USA}

%\collaboration{(ZFOURGE survey collaboration)}

\begin{abstract}
We study galactic star-formation activity as a function of environment and stellar mass over 0.5$<$$z$$<$2.0 using the FourStar Galaxy Evolution (ZFOURGE) survey. We estimate the galaxy environment using a Bayesian-motivated measure of the distance to the third nearest neighbor for galaxies to the stellar mass completeness of our survey, $\log(M/M_\odot)>$ 9 (9.5) at $z$=1.3 (2.0).  This method, when applied to a mock catalog with the photometric-redshift precision ($\sigma_z / (1+z) \lesssim 0.02$) of ZFOURGE, recovers galaxies in low- and high-density environments accurately.  We quantify the environmental quenching efficiency, and show that at $z>$ 0.5 it depends on galaxy stellar mass, demonstrating that the effects of quenching related to (stellar) mass and environment are not separable. In high-density environments, the mass and environmental quenching efficiencies are comparable for massive galaxies ($\log (M/M_\odot)\gtrsim$ 10.5) at all redshifts. For lower mass galaxies ($\log (M/M)_\odot) \lesssim$ 10), the environmental quenching efficiency is very low at $z\gtrsim$ 1.5, but increases rapidly with decreasing redshift. Environmental quenching can account for nearly all quiescent lower mass galaxies ($\log(M/M_\odot) \sim$ 9--10), which appear primarily at $z\lesssim$ 1.0.  The morphologies of lower mass quiescent galaxies are inconsistent with those expected of recently quenched star-forming galaxies.  Some environmental process must transform the morphologies on similar timescales as  the environmental quenching itself. The evolution of the environmental quenching favors models that combine gas starvation (as galaxies become satellites)  with gas exhaustion through star-formation and outflows (``overconsumption''), and additional processes such as galaxy interactions, tidal stripping, and disk fading to account for the morphological differences between the quiescent and star-forming galaxy populations. 
\end{abstract}

\section{Introduction}\label{sec:intro}

The population of galaxies can be broadly classified into two distinct types: quiescent galaxies with
relatively red colors, spheroid-dominated morphologies, and little to no on-going star-formation activity (with star-formation rates [SFRs] much less than their past averages);  and  star-forming galaxies with relatively blue colors, disk-dominated morphologies, and SFRs comparable to (or above) their past averages \citep[e.g.,][]{Strateva2001,Kauffmann2003,Baldry2004,Baldry2006,Bell2008,vanDokkum2011,Schawinski2014}.  In the local universe, it is well-known that these types of galaxies are related to the density of galaxies (the galaxy environment).  Quiescent, spheroidal galaxies are preferentially found in dense environments rich with galaxies \citep[e.g.,][]{Oemler1974,Davis1976,Dressler1980,Balogh2004,Hogg2004,Kauffmann2004,Blanton2009,Peng2010,Woo2013}.   
%
%At $z\sim 1$, several studies have observed the role of environment on quenching the star-formation of galaxies, and those galaxies eventually have migrated to the quiescent population, which is preferentially found in dense environment \todo{I'm not sure what this previous sentence is saying. Do we just mean that "Several studies have found a relationship between environmental density and quenching at $z\sim1$"?} 

How this trend with environment manifests and evolves with redshift is  one of the outstanding questions in galaxy evolution.
Multiple studies have found a correlation between environmental density and the quenching of star-formation at $z\sim1$ \citep[e.g.,][]{Cooper2007,Cooper2010, Cucciati2010,Kovac2010,Kovac2014,Muzzin2012,Balogh2016,Allen2016,Morishita2016,Guo2017}.
In addition, there is some observational evidence that the environment (or by proxy, the density of galaxies) correlates with other galaxy properties out to $z\sim2$ \citep[e.g.,][]{Cucciati2006,Tran2010,Chuter2011,Papovich2012,Quadri2012,Fossati2016,Nantais2016}, and possibly to at least $z\sim 3$ \citep{Darvish2016}. 
Developing a further understanding of the physical processes involved in the quenching of star formation clearly requires better observational measurements, and also requires disentangling the observed correlations between SFR, galaxy structure, and environmental density, out to these higher redshifts.
%However, we still have a limited understanding of the physical processes, owing to the environmental cause these apparent correlations between galaxy density and galaxy SFRs. 

In the low redshift Universe ($z \lesssim 0.1$), it has been shown that the respective relationships between stellar mass and environment on quenching are largely separable \citep[e.g.,][]{Baldry2006,vandenBosch2008,Peng2010,Kovac2014}, implying there are two distinct quenching processes at work: one that correlates with stellar mass (independent of environment) and one that correlates with galaxy environment (independent of stellar mass). 
%
%More massive galaxies are likely to become quiescent regardless of their environment, and galaxies in overdense environments are likely to become quiescent regardless of stellar mass.
%
\cite{Peng2010} show the separability of the effects of stellar mass and environment on the quiescent fraction of galaxies in SDSS (at $z < 0.1$) and zCOSMOS ($0.3 < z < 0.6$). Similarly, \cite{Kovac2014} used mass-matched samples of central and satellite galaxies to show that the quiescent fraction of centrals is primarily related to stellar mass and is almost independent of overdensity (environment), indicating that they are mainly quenched by a process related to stellar mass, at least to the stellar mass limit of their data ($\log M/M_\odot > 9.8$ at $z=0.4$). On the other hand, an additional {environmental quenching} process is required to explain the observed quiescent fraction of satellite galaxies, which increases with galaxy overdensity. 

These separable effects of stellar mass and environment on galaxy properties have been observed out to $z\sim3$ \citep[e.g.,][]{Quadri2012,Muzzin2012,Lee2015,Darvish2016}, but have been limited to more massive galaxies, $\log M/M_\odot \gtrsim 10$, due to the depth of the surveys. To this mass limit, these studies also found no evidence that the properties of star-forming galaxies strongly depend on environment:   there is no significant change in the median SFR and specific SFR (for star-forming galaxies) with environment at fixed stellar mass, suggesting that the independence of SFR-mass sequence on environment has been in place at $z\sim3$ \citep[but see][]{Jian2017}. The obvious qualifier is that it has only been possible to study the relatively higher mass galaxies, and it is unknown if these results extend to lower mass galaxies.

\par Physical explanations for cessation of star-formation in galaxies can also be broadly classified into mechanisms that related to mass (halo mass, supermassive black hole mass, or stellar mass) or to environment. A galaxy's halo mass provides a natural quenching mechanism related to mass \citep[e.g.,][]{Rees1977,White1978,Gabor2011,Gabor2015}. It is generally argued that the intra-halo gas in halos above $\sim 10^{12}\Msol$ exists at temperatures high enough \citep{Birnboim2003,Keres2005} to shock-heat infalling gas from the inter-galactic medium at the virial radius, preventing the fueling of star-formation in the galaxies \citep{Dekel2006,Cattaneo2006,Birnboim2007}.  Another quenching mechanism that may be related to galaxy mass is the feedback from an active galactic nucleus (AGN) \citep{Granato2004,Springel2005,Croton2006,Bower2006,Cattaneo2006,Somerville2008,Knobel2015,Terrazas2016}. In contrast, star-formation suppression of galaxies in low-mass halos can be driven by energetic feedback from supernova explosions and stellar winds \citep[e.g.,][]{Larson1974,Dekel1986}.

\par There are also physical processes that operate preferentially in dense environments. One of them is the rapid stripping of cold gas via ram pressure as the galaxy passes through a hot gaseous medium, causing  abrupt quenching \citep{Gunn1972,Abadi1999}.  In contrast, if only the hot gas in the outer parts of galaxies is stripped, the galaxy may continue forming new stars until all fuel is exhausted. Consequently, this ``strangulation'' (also called ``starvation'') results in the gradual decline of star-formation rate \citep{Larson1980,Balogh1997}. Note, however, that both of these gas-stripping processes will primarily modify the color and SFR of a galaxy, without transforming the galaxy morphology \citep[e.g.,][]{Weinmann2006,vandenBosch2008}\footnote{While this is true in morphology as traced by stellar mass, for morphology as traced by light in any passband, even near-IR, the higher luminosity of young stars will make the star-formation disks more prominent and will lead to significant changes in visual appearance of morphology as the star-formation fades \citep[e.g.,][]{Fang2013}. We return to this point about ``disk fading'' in \S~5.}.  Satellite galaxies orbiting within dark matter halos may also be subject to tidal stripping as they experience tidal forces due to the central galaxy, due to other satellite galaxies, and due to the potential of the halo itself \citep[e.g.,][]{Read2006}.  Higher density environments can also lead to enhanced merger rates, which may also affect quenching \citep{Peng2010}. Recently, \cite{McGee2014} pointed out that the gas outflows that are ubiquitous among star-forming galaxies may also affect the quenching of satellites. According to this scenario, which they refer to as ``overconsumption," vigorous star formation in recently-accreted satellites may drive outflows that will exhaust the gas supply in the absence of cosmological accretion. These authors also demonstrate that the timescale for satellite quenching due to overconsumption can be much shorter than the time for the gas to be stripped through dynamical processes.

\par Another clue regarding environmental quenching mechanisms is the observed correlation between the properties of satellites (i.e., specific SFR, colors, and gas fraction) and their more massive central galaxies. The correlation is such that the satellites of quiescent centrals are more likely to be quenched than the satellites of star-forming centrals, even at fixed stellar mass. This phenomenon was originally presented by \cite{Weinmann2006} and is referred to as ``galactic conformity". There is growing observational evidence of galactic conformity in both the local Universe \citep{Kauffmann2010,Kauffmann2013,Knobel2015,Phillips2014,Phillips2015} and out to $z\sim2$, even though the signal is weaker at high redshift  \citep{Hartley2015,Kawinwanichakij2016,Hatfield2016}. 
%
%These authors generally show that satellite quenching is connected to the star-formation properties of the central galaxy as well as to the mass of the halo. \todo{Is it really true that these authors generally find that it happens at fixed halo mass? I actually don't remember what other authors found. We saw a hint only in one redshift range, right?} 
%
Broadly speaking, the different environmental processes discussed here are expected to act with different strengths and over different timescales as a function of galaxy stellar and halo mass.  Therefore, measuring how (stellar) mass and environmental quenching evolve with stellar mass and redshift provides constraints on the quenching processes, particularly at higher redshift ($z \gtrsim 1$), when galaxy specific SFRs are higher.

\par In this paper we primarily focus on how the quenching of galaxies correlates with galaxy stellar mass and environment, and how these evolve with redshift. However, we do not attempt to separate our sample into central or satellite galaxies. Rather, we will denote the environmental density, based on the local overdensity of galaxies compared to the mean, as ($1 + \delta$). We will make use of the deep NIR imaging and high photometric redshift accuracy from the FourStar Galaxy Evolution (ZFOURGE) survey, which allows us to compute accurate estimates of the environment for galaxies to fainter magnitudes and with higher completeness than is possible with either ground-based spectroscopy ($K_{s} < 24$~AB mag, \citealt{Nanayakkara2016}, for emission-line galaxies) or space-based spectroscopy ($JH_{140} < 24$~AB mag, \citealt{Fossati2016}).  In contrast, the ZFOURGE data provide precise photometric redshifts for galaxies to $K_s \simeq 25.5-26$~AB mag, substantially fainter than what is possible with spectroscopy. As a result, the ZFOURGE data allow us to study environmental impact of quenching for galaxies with low stellar mass out to high redshift ($\log (M/M_\odot)$ $\simeq$ 9.5 at $z=2$). 

Because we quantify quenching as a function of both stellar mass and local environmental density, throughout this paper we refer to ``mass quenching" and ``environmental quenching" processes. This does not imply that stellar mass and environmental density directly cause quenching. For instance, black hole mass or central stellar mass density may have a more direct relationship to quenching than stellar mass \citep{Terrazas2016, Woo2015}, but because these quantities correlate with stellar mass, they will result in a measurable "mass quenching" effect. Similarly, the estimator of environmental density that we use may only be correlated with, rather than directly measure, the aspects of a galaxy's location or environment that actually cause quenching. 

\par The outline of this paper is as follows. In Section~\ref{sec:data}, we describe the ZFOURGE catalog and our galaxy sample selection criteria. In Section~\ref{sec:method}, we describe the method for estimating the environmental densities using photometric redshifts, and we validate our method using simulated catalogs from a semi-analytic model (described further in Appendix~\ref{sec:densitysimulation}). In Section~\ref{sec:result}, we discuss how the fraction of quiescent galaxies varies with stellar mass and environment, and we compute from these the quenching efficiency for both variables out to $z=2$.  
%
% we can cut all this, I think:
%We will show that the quiescent fraction for galaxies is higher in higher density environments at fixed stellar mass.  However, this depends on the stellar mass and redshift, and while we see strong environmental quenching of higher mass galaxies ($\log (M/M_\odot) > 10.5$) out to the redshift limit of this study, the environmental quenching for lower-mass galaxies ($\log (M/M_\odot) \simeq 9-10$) decreases beyond $z\sim 0.5$, becoming consistent with no environmental quenching at $z\sim 2$.   
%
In Section~\ref{sec:discussion}, we discuss the relative importance of environmental processes in the buildup of red galaxies in dense environments.  We investigate whether the cause of environmental quenching is indicated in the morphological distribution of lower-mass quiescent galaxies.  In addition, we consider how our results constrain timescales of environmental quenching and therefore the physical processes responsible. In Section~\ref{sec:summary}, we present our summary. Throughout, we adopt the following cosmological parameters where appropriate, $H_0 = 70 \unit{km\;s^{-1}\;Mpc^{-1}}$, $\Omega_{m} =0.3$, and $\Omega_{\Lambda} = 0.7$. All magnitudes are expressed in the AB system.

\section{Data and Sample Selection}
\label{sec:data}

We select galaxies at $0.5 < z < 2.0$ from the ZFOURGE survey \citep{Straatman2016}.  The survey is composed of three $11' \times 11'$
fields with coverage in regions of the CDFS \citep{Giacconi2002}, COSMOS \citep{Scoville2007}, 
and UDS \citep{Lawrence2007} that overlap with the Cosmic Assembly Near-IR Deep Extragalactic Legacy Survey \citep[CANDELS][]{Grogin2011,Koekemoer2011}, which also provide \textit{Hubble} Space Telescope, high-angular resolution imaging for 0.6--1.6~$\mu$m \citep[see, e.g.,][]{VanderWel2012}. The ZFOURGE medium--band near-IR imaging reaches depths of $\sim 26$ AB mag
in $J_{1}, J_{2}, J_{3}$ and $\sim 25$ AB mag in $H_{s}, H_{l}$ and includes the vast amount of deep, multiwavelength imaging available in these legacy fields.  The ZFOURGE catalogs are complete for galaxies to $K_{s} \simeq 25.5-26.0$~AB mag \citep[see][]{Straatman2016}.

%\vspace{1cm}
\subsection{Photometric Redshifts}

\par The ZFOURGE catalogs include photometric redshifts and rest-frame colors calculated using EAZY \citep{Brammer2008} from the $0.3-8$~$\mu$m photometry for each galaxy.  Of import here, ZFOURGE uses templates and photometric zeropoints that are iteratively adjusted during the fitting process to improve accuracy of the photometric redshifts.  

\par The precision of photometric redshifts has the ability to potentially introduce  spurious signals or wash out structure \citep{Cooper2005,Quadri2012}.  However, our estimates of the quality of the ZFOURGE photometric redshifts show them to be very accurate, and as we demonstrate below, sufficient to recover galaxy environmental densities.  By comparing photometric redshifts of galaxy pairs and to spectroscopic subsamples, \cite{Straatman2016} show that the typical photometric redshift uncertainties are $\sigma_z/(1+z)$= \editone{0.01--0.02} to the $K_s$-band magnitude limit for galaxies between $z = 0.5$ and $z=2.0$, with negligible dependence on galaxy color, \editone{but there is dependence on magnitude and redshift \citep[see their section 5.4]{Straatman2016}}. Quantitatively, at $K_s < 25.0$ AB mag photometric redshift uncertainties of quiescent galaxies at $z=2$ are better than that of star-forming galaxies at the same redshift only about 5\%.  Other studies with ZFOURGE have shown that these redshifts are sufficient to identify protoclusters out to $z\sim 2$ \citep[e.g.,][]{Spitler2012,Yuan2014,Forrest2017}. 

In addition to the photometric redshifts, we also make use of stellar masses for galaxies provided in the ZFOURGE catalogs.  The stellar masses were derived by fitting stellar population models to the photometry using FAST
\citep{Kriek2009}, assuming exponentially declining star formation
histories, solar metallicity, and a \cite{Chabrier2003} initial mass function.
\vspace{0.5cm}

\startlongtable
\begin{deluxetable}{cc|cc}
 \tablecaption{ Stellar mass completeness limits for ZFOURGE galaxies at $0.5 < z < 2.5$
 \label{table:masscompletetable}}
 \tablehead{
%  \multirow{2}{*}{Redshift} \vspace{0.1cm}&  \colhead{$\log(M(z))$}  \\ \\
  \colhead{Redshift} \vspace{0.1cm}&  \colhead{$\log(M / \mathrm{M}_\odot)$}  & 
 \colhead{Redshift} \vspace{0.1cm}&  \colhead{$\log(M / \mathrm{M}_\odot)$}    
  } 
 \startdata 
%\vspace{0.2cm}
%0.3&\phn7.6\\
% 0.4&\phn7.9\\
%0.5& \phn8.09 &  1.3 & \phn 9.01 \\
%0.6&\phn8.25 & 1.4 & \phn 9.09\\
%0.7&\phn8.40 & 1.5 & \phn9.17 \\
%0.8&\phn8.52&  1.6 & \phn 9.24 \\
%0.9& \phn8.64 & 1.7 & \phn 9.31 \\
%1.0&\phn8.74 & 1.8 & \phn 9.38\\
%1.1&\phn8.84 & 1.9 &\phn 9.44 \\
%1.2& \phn8.93 &  2.0 &\phn 9.51
0.5& \phn8.09  & 1.6 & \phn 9.24 \\
0.6&\phn8.25   & 1.7 & \phn 9.31 \\
0.7&\phn8.40   & 1.8 & \phn 9.38\\
0.8&\phn8.52   & 1.9 &\phn 9.44 \\
0.9& \phn8.64  & 2.0 &\phn 9.51 \\
1.0&\phn8.74   & 2.1 &\phn 9.57 \\
1.1&\phn8.84   & 2.2 &\phn 9.62 \\
1.2& \phn8.93  & 2.3 &\phn 9.68 \\ 
1.3& \phn 9.01 & 2.4 &\phn 9.73 \\
1.4 & \phn 9.09 & 2.5 &\phn 9.79 \\
1.5 & \phn 9.17 &     &   
\enddata
%\tablenotetext{a}{Adjusted for inflation}
%\tablenotetext{b}{Accounts for the change from page charges to digital quanta in April, 2011}
%\tablecomments{Note that {\tt \string \colnumbers} does not work with the vertical line alignment token. If you want vertical lines in the headers you can not use this command at this time.}
\end{deluxetable}

\subsection {Stellar Mass Completeness}

\label{sec:masscomplete}
\par 
%
%The stellar mass-completeness limit is crucial for our analysis. 
%
Because we are concerned with the galaxy quiescent fractions, it is important that we use a dataset that is complete in stellar mass for both star-forming and quiescent galaxies. 
Quiescent galaxies have higher mass-to-light ratios and therefore will have a higher-mass completeness limit than star-forming galaxies at fixed magnitude.  
Here, we adopt 90\% mass-completeness limits for galaxies with a quiescent stellar population using the technique described by \cite{Quadri2012}. In a given narrow redshift bin, we select all quiescent galaxies and scale their fluxes (and therefore their stellar masses) downward until they have the same magnitude as the measured magnitude limit, $K_s$ = 25.5~AB mag, for all three ZFOURGE fields. Then we define the mass-completeness limit as the stellar mass at which we recover 90\% of the dimmed galaxies at each redshift. We provide the adopted completeness limits for ZFOURGE at $0.5 < z < 2.5$ in Table~\ref{table:masscompletetable}.

\subsection{Selection of Quiescent and Star-forming Galaxies}
\label{sec:samplesel}

\par Our goal is to measure the fraction of quiescent galaxies as a function of stellar mass, environment, and redshift.  From the parent sample of all galaxies in the ZFOURGE catalog, we first select all well-detected galaxies (\texttt{USE} flag = 1) and group them into three bins of redshift,  $0.5 < z < 1.0$, $1.0 < z < 1.5$, and $1.5 < z < 2.0$.  We then further subdivide the samples into bins of galaxy stellar mass, $8.8 <\log (M/\msol) < 9.8$, $9.8 < \log (M/\msol) < 10.5$, and $10.5 < \log (M/\msol) < 11.5$.   In each of these bins, we classify galaxies as star-forming or quiescent using their rest-frame $U-V$ and $V-J$ colors, denoted by $(U-V)_0$ and $(V-J)_0$, respectively. This $UVJ$ color--color space is useful to separate galaxies with colors of quiescent and star-forming stellar populations \citep[including the affects of dust attenuation;][]{Williams2009,Patel2012}. Due to the small systematic variations in the rest-frame colors of galaxies at fixed stellar mass and redshift in different surveys, we follow our previous method \citep[\editone{see their section 2.2}]{Kawinwanichakij2016} to self-calibrate the region in the $UVJ$ color--color space to  delineate star-forming from quiescent galaxies.  We then select quiescent galaxies whose rest-frame colors satisfy,
 \begin{eqnarray} \label{eq:uvj}
  (U-V)_{0} &>& 1.2 \times (V-J)_{0} + 0.2   \nonumber \\
  (U-V)_{0} &>& 1.3  \\
   (V-J)_{0} &<&1.6. \nonumber
\end{eqnarray}
A summary of the number of galaxies from each galaxy mass, redshift, and density subsample is given in Table~\ref{table:sampleno}.

\startlongtable
\begin{deluxetable*}{cccccccc}
\tabletypesize{\footnotesize}
\tablecolumns{2} 
\tablewidth{0pt}
 \tablecaption{Number of quiescent and star-forming galaxies in different stellar mass and density quartile in ZFOURGE at $0.5 < z < 2.0$
 \label{table:sampleno}}
 \tablehead{
 \multirow{2}{*}{Stellar Mass Range} \vspace{-0.1cm}&  \multirow{2}{*}{Redshift Range} & \multicolumn{3}{c}{Lowest-Density Quartile}  & \multicolumn{3}{c}{ Highest-Density quartile} \\ 
  \colhead{} \vspace{-0.1cm}& \colhead{} & \colhead{$N_{\mathrm{Quiescent}}$} & \colhead{$N_{\mathrm{Star-forming}}$} & \colhead{$N_{\mathrm{Total}}$} & \colhead{$N_{\mathrm{Quiescent}}$}  & \colhead{$N_{\mathrm{Star-forming}}$} & \colhead{$N_{\mathrm{Total}}$}\\
\vspace{0.1cm}} 
 \startdata 
%\vspace{0.2cm}
 $8.8 < \log(M/\Msol) < 9.8$ & $ 0.5 < z < 1.0 $ &  24 &         662 &         686 &         113 &         427 &         540\\
                          & $ 1.0 < z < 1.5$  &    10 &         491 &         501 &          22 &         406 &         428 \\
                          & $ 1.5 < z < 2.0$ &  8 &         259 &         267 &           5 &         212 &         217 \\ \\
 \cline{1-8} 
                          &                  &             &              &     &       \\
$9.8 < \log(M/\Msol) < 10.5$&$ 0.5 < z < 1.0 $&  39 &         129 &         168 &          96 &         138 &         234 \\
                          & $ 1.0 < z < 1.5$&   32 &         146 &         178 &          42 &         144 &         186 \\
                           &  $ 1.5 < z < 2.0$ &  24 &         196 &         220 &          40 &         165 &         205\\ \\
\cline{1-8}
                       &                  &             &              &     &       \\
$10.5 < \log(M/\Msol) < 11.5$   & $ 0.5 < z < 1.0 $ &  30 &          33 &          63 &          87 &          50 &         137\\ 
                              &  $ 1.0 < z < 1.5$  &  15 &          37 &          52 &          51 &          51 &         102\\ 
                              &  $ 1.5 < z < 2.0$ &  19 &          64 &          83 &          54 &          74 &         128
 \enddata
%\vspace{-0.1cm}
%see /Users/lalitwadee/ZF_environ/nearest_neighbor/plotuvj.pro.pro
\end{deluxetable*}

\begin{figure}
\epsscale{1.2}
\plotone{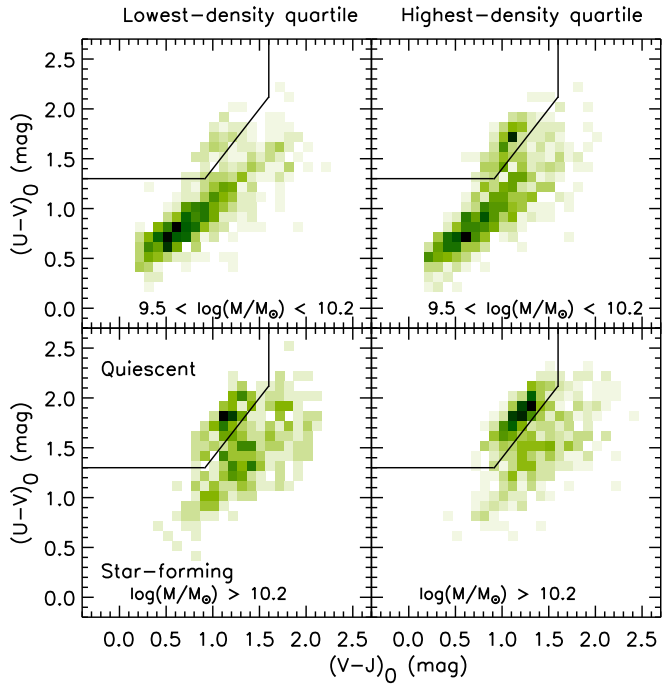}
%see /Users/lalitwadee/ZF_environ/nearest_neighbor/newstat/plotuvj_density.pro

%see /Users/lalitwadee/ZF_environ/nearest_neighbor/plotuvj_density.pro
\caption{Rest-frame $U-V$ versus rest-frame $V - J$ color for galaxies at $0.5 < z < 2.0$ in the lowest-density quartile (left panels) and the highest-density quartile (right panels),
with lower stellar masses \editone{$9.5 < \log (M/\Msol) < 10.2$} (top panels) and higher stellar masses $\log (M/\Msol) > 10.2$  (bottom panels). In each panel, the darkness of the shading is proportional to the number of galaxies in that region.   Galaxies in the  upper left region of each plot (separated by the solid line) have colors of quiescent stellar populations; galaxies outside this region have colors of star-forming populations combined with dust attenuation \citep[see][]{Williams2009}. } 

\label{fig:uvjdiagram}
\end{figure}

Figure~\ref{fig:uvjdiagram} shows the $UVJ$ color--color diagram for galaxies in our ZFOURGE samples  at $0.5 < z < 2.0$ split into low stellar mass ($8.8 < \log(M/\Msol) < 10.2$ and high stellar mass ($10.2 < \log(M/\Msol) < 11.5$) subsamples, and as a function of environment as defined by local overdensity (see Section~\ref{sec:method} below). While all panels show a similar range of galaxy colors, differences in the distributions are clearly evident with mass (and also environment).  The distribution of lower-mass galaxies is weighted more toward bluer star-forming objects (by number), while the distribution of higher mass star-forming galaxies show higher dust attenuation, consistent with other studies \citep[e.g.,][]{Wuyts2011}. All panels also show the existence of a ``red sequence'' of quiescent galaxies, but this is much more pronounced in the higher density regions (denser environments), which we discuss more below.  

%\vspace{1cm}
\subsection{Structural Morphological Parameters}
\label{sec:morphmethod}
In our analysis, we study the morphological differences between quiescent and star-forming galaxies in different environments and as a function of stellar mass.   The majority of galaxies in our sample fall within the CANDELS coverage from HST/WFC3, with effective semi-major axis, $a_\mathrm{eff}$,  and S\'{e}rsic index, $n$, measured by \citet{VanderWel2012} using the HST/WFC3 F160W ($H_{160}$)--band imaging. We refer the reader to \citeauthor{VanderWel2012} for the measurement of random and systematic uncertainties of the estimated morphological parameters using simulated galaxy images. We cross--matched the sources in our catalog with those of \citeauthor{VanderWel2012} \editone{The fractions of our galaxy sample with available morphological parameters from \citeauthor{VanderWel2012} are 80\%, 90\%, and 82\% for CDFS, COSMOS, UDS, respectively. We note that there are 10\%-20\% of our galaxy sample that have no morphological information from HST/WFC3 because those galaxies are in the regions around the edges of ZFOURGE fields where there is no HST/WFC3 coverage.} We further define the circularized effective radius as $r_\mathrm{eff} = a_\mathrm{eff} \sqrt{q}$, where $a_\mathrm{eff}$ is the effective semi-major axis and $q = b/a$ is the ratio of the semi-minor to semi-major axis. In addition to the morphological parameters from \citeauthor{VanderWel2012}, we calculate the stellar mass surface density inner 1kpc which we describe the procedure in Appendix~\ref{sec:morph}. 

\section{Measurement of Galaxy Density as Estimate of Environment}
\label{sec:method}

In this work we estimate the local galaxy (projected) overdensity
using the distance to the $N$th nearest neighbor, $d_N$.  This distance has often been used as a measure of the overdensity, with $N$ 
typically varying from 3 to 10 \citep[e.g.,][] {Dressler1980,Baldry2006,Muldrew2012}. 
We then are able to define the environment of a galaxy in terms of the dimensionless  overdensity, $1+\delta$,  defined as
 \begin{equation} \label{eq:overdensity}
 (1+\delta)_{N} = 1 + \frac{\Sigma_N - \left \langle \Sigma \right \rangle}{\left \langle \Sigma  \right \rangle},
\end{equation}

 %\begin{equation} %\label{eq:overdensity}
 %(1+\delta)_{N} = 1 + \frac{\Sigma_N %- \left \langle \Sigma_N  \right %\rangle}{\left \langle \Sigma_N  %\right \rangle},
%\end{equation}
%
where $\Sigma_N=N/(\pi d_N^2)$, is the local surface density of a galaxy based on the distance to the $N$th nearest neighbor and $\left \langle \Sigma  \right \rangle$ is the average surface number density of galaxies over the whole field. We then take $(1+\delta)$ to denote the fractional density of galaxies with respect to the mean (as a function of redshift).  

\par We improve our measurement of overdensity using an estimator for the $N$th nearest neighbor introduced by \cite{Ivezic2005}. The distances to all $N$ nearest neighbors provide the information about local density (overdensity) of galaxies. Motivated by Bayesian probability framework, we incorporate the projected distance to $N$th nearest neighbors into density estimator, and we additionally take into account information from the projected distances to the first $N-1$ nearest neighbors.  \citeauthor{Ivezic2005} demonstrate that this increases the accuracy of the overdensity compared to the traditional $N$th nearest neighbor metric because it uses the distances to the 1st, 2nd, ... $N$th neighbors and is less subject to projection effects (\editone{see Appendix B of Ivezic et al.\ 2005}).  One of the advantages of this estimator is that it provides a good estimate of the ``local density,'' which corresponds to scales internal to galaxy group halos,  provided $N$ is relatively small \citep[see][]{Muldrew2012}. As many of the environmental trends we find appear to correlate with group--sized halos, we adopt $N$=3 for the analysis here.    Specifically, we use the estimator as given by \cite{Cowan2008}:
\begin{equation}
\label{eq:bayesian}
 \Sigma^{\prime}_N=C \frac{1}{\Sigma_{i=1}^{N} d_{i}^2}.
\end{equation}
Here, we take the third nearest neighbor (3NN) distance, where we empirically determine the constant, $C$,  by requiring that $\left \langle \Sigma^{\prime}_N  \right \rangle$ be equal to that for a uniform density of galaxies with the same total number and area as in our ZFOURGE dataset.

\begin{figure}
\epsscale{1.0}
%\plotone{overdensity_redshift}
\plotone{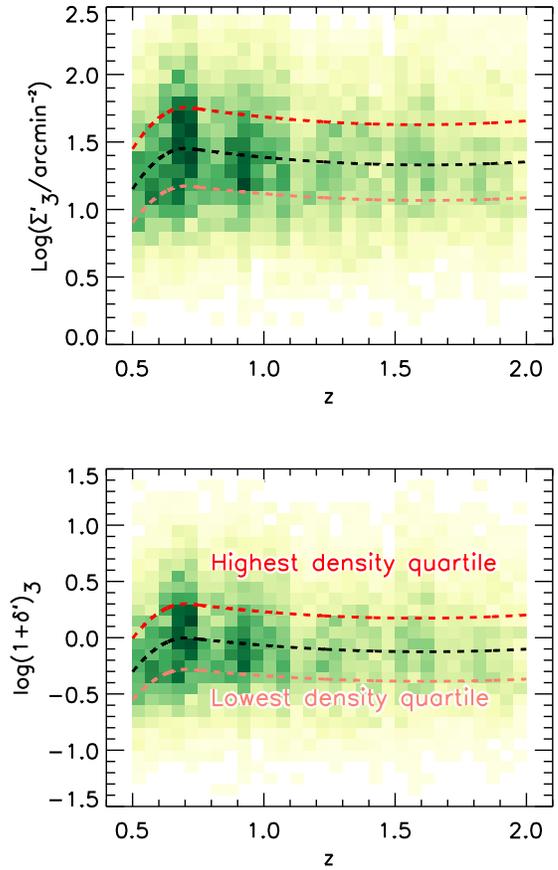}
%see /Users/lalitwadee/ZF_environ/nearest_neighbor/newstat/calcsigma_1pd_ver2.pro
%see /Users/lalitwadee/ZF_environ/nearest_neighbor/density_quartile_ver2.pro for the older version (bayesian_n3_z_240517.ps)
\caption{\emph{Top:} The projected density computed from the Bayesian 3rd nearest neighbor ($\Sigma^{\prime}_3$) of each galaxy in three combined ZFOURGE fields as a function of redshift. \editone{The black, light-red, and red dashed lines show the median, bottom, and top quartile (25th percentiles) of the distribution, derived from a spline quartile regression applied to the data (see text)}. \emph{Bottom}: The corresponding overdensity of ZFOURGE galaxies as a function of redshift computed using the same method. We again show the median, \editone{bottom, and top quartiles}. We define galaxies in the upper and lower quartiles of overdensity to be in ``high'' and ``low'' density environments, respectively. In each panel, the darkness of the shading is proportional to the number of galaxies in that region.} 
\label{fig:overdensity}
\end{figure}
%Note that the overdensity increases with increasing redshift due to the decreasing of average surface number density of galaxies with redshift. 

\par For this study, we calculate the 3NN distance, $\Sigma^{\prime}_3$, for each galaxy in the ZFOURGE catalog. At each redshift, we consider all galaxies more massive than the mass completeness limits given in Table~\ref{table:masscompletetable}.  For each galaxy, we measure the density only considering galaxies with a photometric redshift separation that is  2.5 times the estimated redshift uncertainty i.e., $2.5\times0.02(1+z_{\mathrm{phot}})$\editone{, where we adopt the factor of 0.02 as a conservative redshift uncertainty.}  We then compute the overdensity of galaxy, $\log(1+\delta^{\prime})_3$, using the Bayesian-motivated estimate of the local surface density of galaxies derived from the 3rd nearest neighbor by substituting $\Sigma^{\prime}_3$ into $\Sigma_N$ of Equation~\ref{eq:overdensity}. In Appendix~\ref{sec:densitysimulation}, we verify using a mock galaxy catalog that this method recovers well the true projected overdensity for data with the photometric accuracy of ZFOURGE in that it faithfully recovers galaxies in the highest and lowest density quartiles with a minimal amount of contamination. Our tests also showed that $N$=3 provides a good compromise between the
accuracy of measured galaxy overdensity and the ability to probe group-sized environments, which is appropriate for our study here. However, we experimented using $\Sigma^\prime_N$ with $N$=2,5,7 and find that our main conclusions are unaffected by the choice of $N$.

\begin{figure*}
\epsscale{1.15}
\plotone{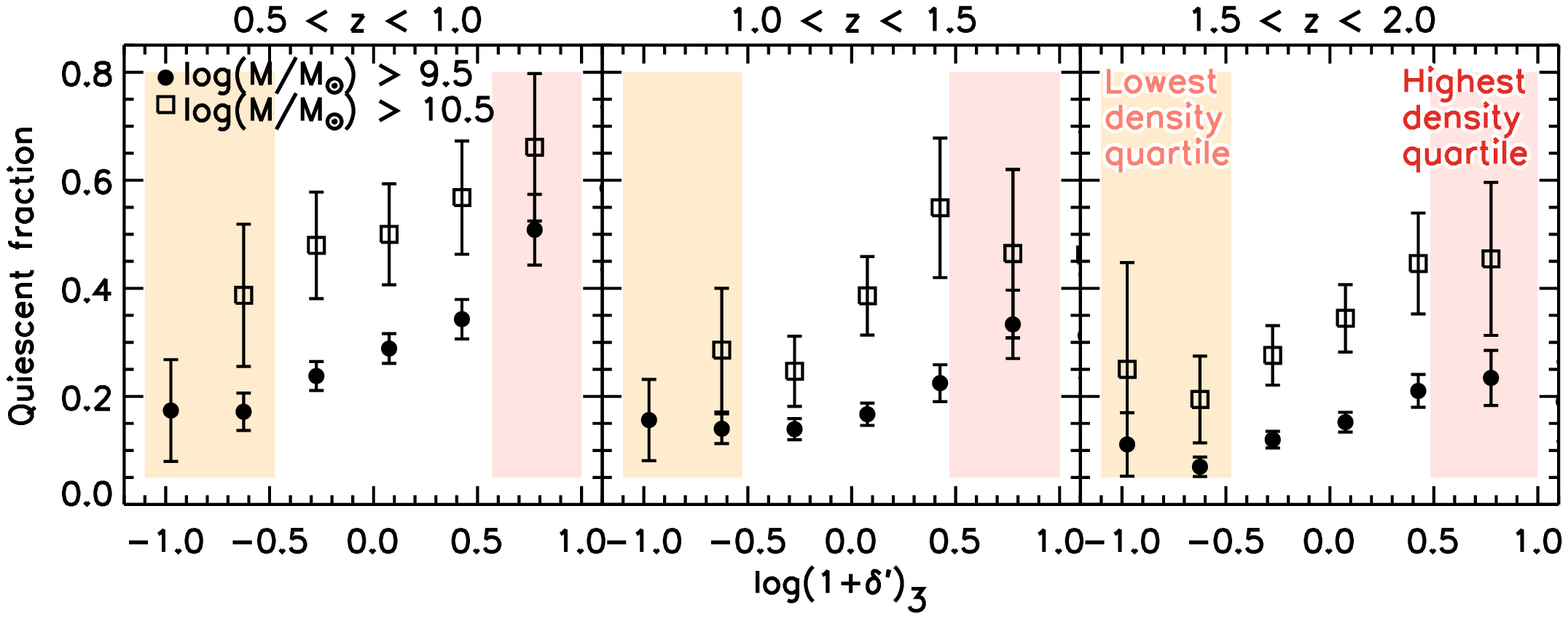}
\caption{Quiescent fraction versus overdensity in three different redshift ranges for two mass-selected samples. The quiescent fractions of galaxies are determined in bins of 0.4 dex of $\log(1+\delta^{\prime})$. The error bars indicate $1\sigma$ uncertainties based on Poisson statistics. The shaded regions in each panel indicate the lower and upper 25th percentiles of $\log(1+\delta^{\prime})$, where we define low- and high-density environments.}
% see /Users/lalitwadee/ZF_environ/nearest_neighbor/qfrac_delta.pro
% see /Users/lalitwadee/ZF_environ/nearest_neighbor/newstat/qfrac_delta.pro
\label{fig:qfracoverdensity}
\end{figure*}

In addition, we have tested for ``edge effects'' by excluding galaxies  from our analysis that are $\sim20$ arcseconds  (larger than the 3NN distances for the galaxies in the lowest-density quartile) from the survey edges. \editone{This does not affect the main results for the differences in the quiescent fractions as a function of stellar mass and environment that are presented in Section~\ref{sec:result}.} To be even more conservative, we then tested for edge effects by excluding galaxies that are two times that distance from the survey edges ($\sim40$ arcseconds), and find no change in our results,  although the uncertainties in quiescent fractions increase as the sample size decreases.   We therefore apply no correction for the edge effects in this study.   

\par Figure~\ref{fig:overdensity} shows the projected density, $\Sigma^{\prime}_3$, and overdensity, $\log(1+\delta^{\prime})$,
computed from the 3NN ($\Sigma^{\prime}_3$) of each ZFOURGE galaxy as a function of redshift. We calculated the median, \editone{bottom and top quartiles of the distribution (i.e., the bottom and top 25th percentiles)} to determine the relative overdensity of each galaxy. \editone{We determined these quartiles using a spline linear regression implemented with the COnstrained B-Splines (cobs) package in R \citep{cobs,Feigelson}.}   We find that, over the redshift range 0.5 -- 2.0, the projected density ($\Sigma^{\prime}_3$) at the lower and upper quartiles of overdensity are about 13 and 43 galaxies per arcmin$^2$, respectively. In the following analysis, we define a galaxy to be in a low (high)-density environment if it has a  overdensity $\log(1+\delta^{\prime})$ less (greater) than the lower (upper)
25th percentile. We will interchangeably use  the terms low/high-density environments (hereafter $\delta_{25}$/$\delta_{75}$) with the lowest/the highest-density quartiles. 

\section{Results}
\label{sec:result}
%\section{Evolution of Quiescent Fraction with Environment, stellar mass, and redshift}
\label{sec:qfrac}

  In this section we calculate the quiescent fraction as a function of stellar mass and environment, using the overdensities  ($1+\delta^{\prime}$) derived above. We use these fractions to estimate the environmental quenching efficiency and (stellar) mass quenching efficiency as defined below. 
%
%Then we investigate the dependence of average quiescent fraction of galaxies and quenching efficiencies on overdensity, stellar mass, and redshift. 

\subsection{Evolution of Quiescent Fraction with Environment and Redshift}

\begin{figure*}
\epsscale{0.75} % CJP decided this fig could be a little smaller - less emphasis.
\plotone{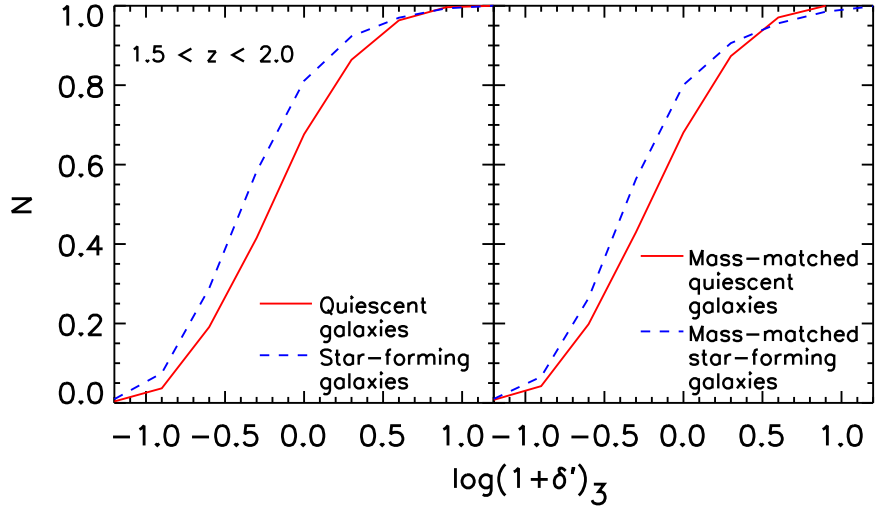}
\caption{\emph{Left:} Cumulative distribution of overdensities $\log(1+\delta^{\prime})$ for quiescent galaxies (red solid line) and star-forming galaxies (blue dash lines) with $\log (M/\Msol) > 9.5$ at $1.5 < z < 2.0$ . A K-S test indicates that we can reject the null hypothesis that these two distributions are drawn from the same parent distribution \editone{with a p-value of $\ll 10^{-3}$}  \emph{Right:} Same as the left panel but for mass-matched and redshift-matched
samples of quiescent and star-forming galaxies. A K-S test indicates the differences in the distributions persists \editone{with a p-value of $\ll 10^{-3}$.}}
% see /Users/lalitwadee/ZF_environ/nearest_neighbor/match_sample_wht.pro
\label{fig:cumuoverdensity}
\end{figure*}
%(equivalent to $>5\sigma$ significance for Gaussian statistics)
% (equivalent to $\sim4\sigma$ significance for Gaussian statistics.)
\par We show the quiescent fraction of galaxies as a function of the overdensity in Figure~\ref{fig:qfracoverdensity}. 
We apply a mass limit of $\log(M/\Msol) > 9.5$ in all three redshift bins, which corresponds to our completeness limit at $z = 2.0$. We also compare our quiescent fraction with those from \cite{Quadri2012} who used the galaxy sample from the UKIDSS Ultra-Deep Survey, so we apply a mass limit of $\log(M/\Msol) > 10.2$, corresponds to completeness limit at $z=2.0$ used by \cite{Quadri2012}. The error bars indicate the 1$\sigma$ uncertainty based on Poisson statistics for the number of quiescent galaxies in a bin. At all redshift ranges, we see evidence for a higher quiescent fraction of galaxies at higher densities. This effect is very strong at $z < 1$ (left panel of Figure~\ref{fig:qfracoverdensity}), but decreases at higher redshift, $z >1$ (middle and right hand panels of Figure~\ref{fig:qfracoverdensity}).
%
%, for lower-mass galaxies  ($\log(M/\Msol) \sim 9.5$). 
%\todo{I'm not too sure if we should say here that it diminishes for low-mass galaxies, since that actually isn't completely obvious in the mass ranges shown in Figure 3, and it is kind of an ambiguous statement anyway. For, instance, the fractional difference between HDE and LDE seems kind of the same in the lowest and highest redshift bins in Figure 3. CJP: Yup, I agree.  Fig 3 just shows redshift--overdensity  relation (no stellar mass)} 
%
This is in agreement with previous studies of star-formation-density relation \citep[e.g.][]{Quadri2012}, where with the ZFOURGE data we have extended the result to lower masses and higher redshifts \citep[for the recent study of galaxy sample with comparable stellar mass and redshift range to our sample see][]{Guo2017}.

\begin{figure*}
\epsscale{1.0}
\plotone{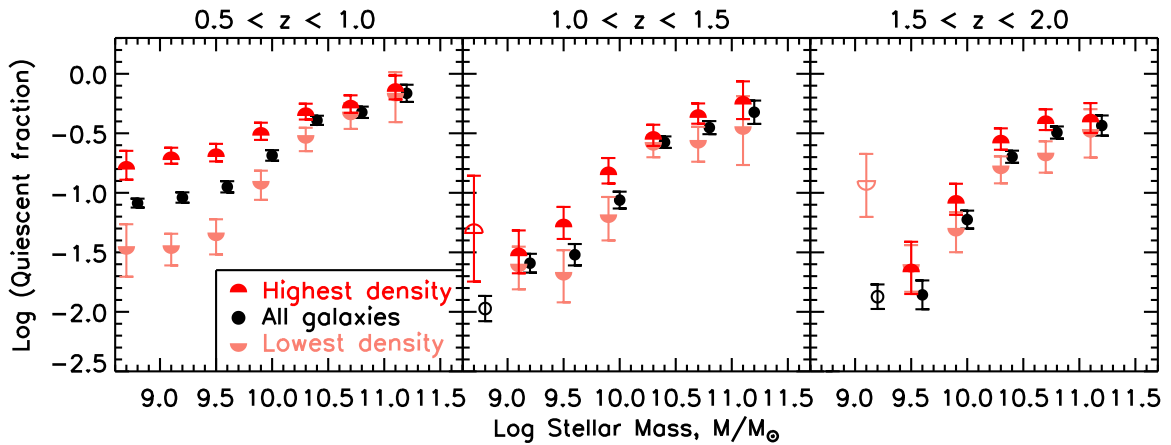}
\caption{Quiescent fraction versus stellar mass in three different redshift ranges for all galaxies (black circles), galaxies in the highest-density quartile (red upper-half circles), and 
galaxies in the lowest-density quartile (light-red lower-half circles). The quiescent fractions are determined in a bin of 0.4 dex in $\log(M/\Msol)$. Open symbols correspond to data below each subsample's respective mass-completeness limit. The error bars indicate $1\sigma$ uncertainties based on Poisson statistics. The quiescent fraction of all galaxies (black circles) are slightly offset along the abscissa for clarity.
Galaxies in denser environments have a higher quiescent fraction in all stellar mass bins out to $z\sim 2.0$, except possibly for the lowest-mass galaxies ($\log(M/\Msol) \simeq 9-10$) at $1.5 < z <2.0$.}
% see /Users/lalitwadee/ZF_environ/nearest_neighbor/qfrac_mass_bayes_ver3.pro
% see /Users/lalitwadee/ZF_environ/nearest_neighbor/newstat/qfrac_mass_bayes_ver3.pro

\label{fig:qfracmass}
\end{figure*}

\par \editone{In this section we have shown that the quiescent fraction of galaxies is higher in denser environments over all redshifts probed in this study. In principle it is possible that this result is caused by differences in the stellar mass distribution and/or the redshift distribution of quiescent and star-forming galaxies. To check for this, we create samples of quiescent and star-forming galaxies such that their stellar mass and redshift distributions are matched following the method described by \cite{Kawinwanichakij2016} (see their section 3.2). The left panel of Figure 4 shows the differences in overdensity between the quiescent and star-forming galaxies before this matching at $1.5<z<2.0$ (a p-value of the differences as measured by a K-S test of $\ll 10^{-3}$) and the right panel shows that the difference persists even after matching the stellar mass and redshift distributions (a p-value of $\ll 10^{-3}$). We obtain even more significant results in our other (lower) redshift bins. We conclude that, at all redshifts  studied  here,  quiescent  galaxies  are  more common in overdense regions compared to star-forming galaxies even taking into account differences in redshift and stellar mass}.

\begin{figure*}
\epsscale{1.0}
\plotone{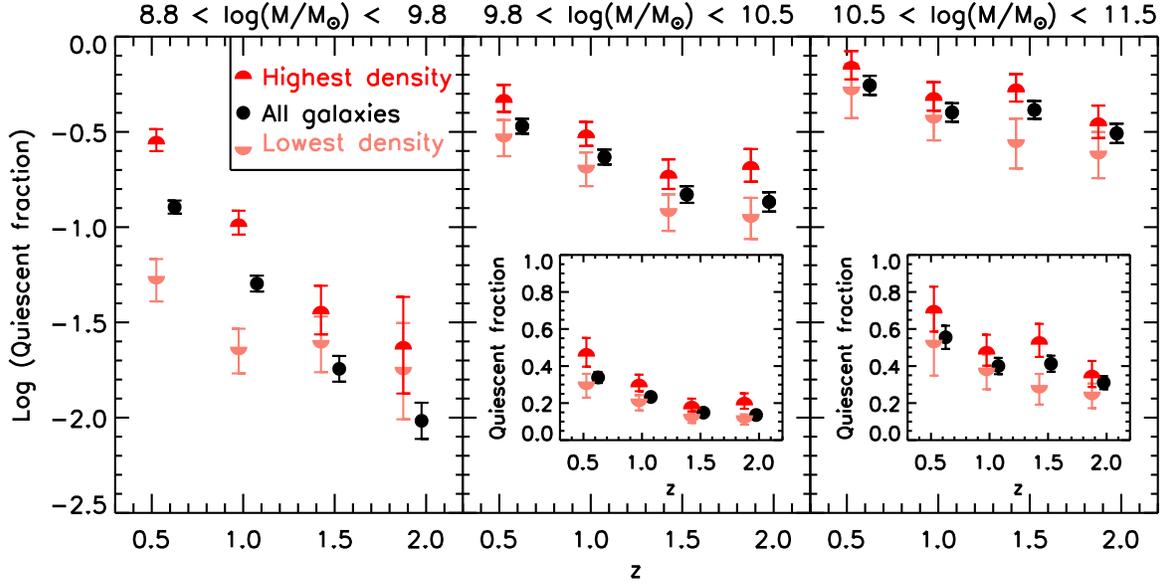}
%\plotone{qfrac_z_N3_Bayes_080617}
\caption{Quiescent fraction versus redshift in three different stellar mass ranges for all galaxies (black circles), galaxies in the highest-density quartile (red upper-half circles), and  galaxies in the lowest-density quartile (light-red lower-half circles). The quiescent fraction of galaxies is determined in bins of $\Delta z= 0.45$, chosen as a balance between redshift sampling and having sufficient statistics in each bin. The error bars indicate $1\sigma$ uncertainties based on Poisson statistics. The quiescent fraction of all galaxies (black circles) are slightly offset along the abscissa for clarity. We find a higher quiescent fraction in denser environments for all stellar mass bins out to $z\sim 2.0$, except for lowest-mass galaxies at $z>1.5$. Quiescent fractions are shown on log scale to present relative quiescent fractions in each stellar mass range.  However, the inset plots in the higher redshift panels show a linear scaling for clarity.}
% see /Users/lalitwadee/ZF_environ/nearest_neighbor/qfrac_z_bayes_ver3.pro

%or newstat/qfac_z_bayes_ver4.pro for inset plot
\label{fig:qfracredshift}
\end{figure*}

\subsection{Evolution of Quiescent Fraction with Stellar Mass and Redshift}

%\begin{figure*}
%\epsscale{1.0}
%\plotone{qfrac_lmass_linearyaxis}
%\caption{Same as Figure~\ref{fig:qfracmass} but the y-axes are in linear scale.}
% see /Users/lalitwadee/ZF_environ/nearest_neighbor/qfrac_mass_bayes_ver2.pro
%\label{fig:qfracmasslinear}
%\end{figure*}

\par Figure~\ref{fig:qfracmass} shows the quiescent fraction as a function of stellar mass in bins of redshift, separating galaxies in the highest ($\delta_{75}$) and lowest ($\delta_{25}$) density quartiles. Qualitatively, at $0.5 < z < 1.5$ galaxies in the highest-density quartile show higher quiescent fractions than galaxies with the same mass in the lowest-density quartile in all stellar mass bins. This is in agreement with \cite{Allen2016}, who show that the fraction of quiescent galaxies with $\log(M/\Msol) \gtrsim 9.5$ at $z\sim0.95$ increases with decreasing distance to the cluster core. At higher redshift, $1.5 < z < 2.0$, this trend persists for more massive galaxies ($\log(M/\Msol) \gtrsim 10$). 

%To summarize, we find that quiescent fractions of galaxies in the highest-density environment are higher than those in the lowest-density environment, and we observe this trend out $z\sim2$.

%\begin{figure*}
%\epsscale{1.0}
%\plotone{qfrac_z_N3_Bayes_linearyaxis}
%\caption{Same as %Figure~\ref{fig:qfracredshift} but the y-axes are in linear scale.}
% see /Users/lalitwadee/ZF_environ/nearest_neighbor/qfrac_z_bayes_ver2.pro
%\label{fig:qfracredshiftlinearyaxis}
%\end{figure*}

\begin{figure*}
\epsscale{1.}
\plotone{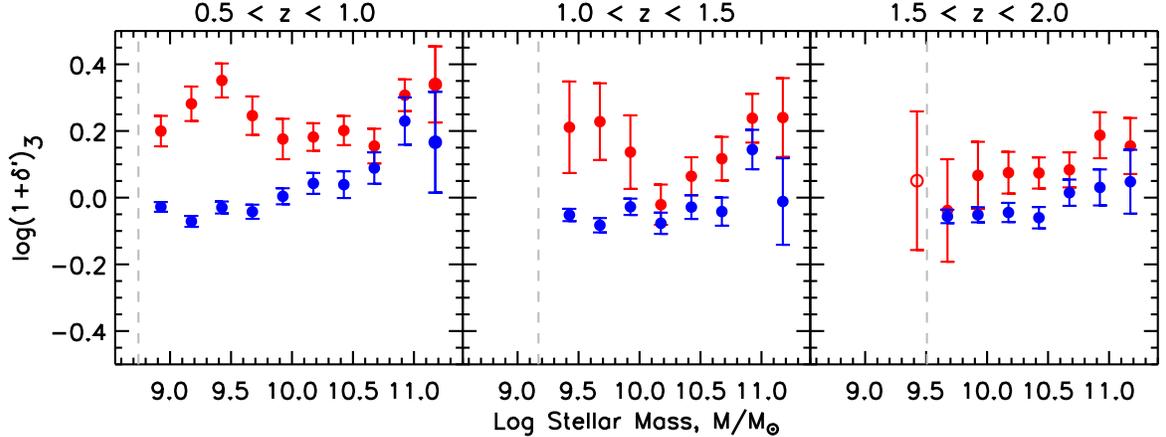}
\caption{Average overdensity versus stellar mass for quiescent galaxies (red circles) and star-forming galaxies (blue circles) in 0.25 dex mass bins. The mass-completeness at each redshift bin is shown as the vertical dashed line. The error bars are the standard deviation of the mean. Even at fixed mass, the quiescent galaxies tend to have higher overdensities than the star-forming galaxies.}
% see /Users/lalitwadee/ZF_environ/nearest_neighbor/delta_mass.pro
% see /Users/lalitwadee/ZF_environ/nearest_neighbor/newstat/delta_mass.pro
\label{fig:overdensitymass}
\end{figure*}

%There is also no significant evolution of average overdensity with redshift for star-forming galaxies.
\par 
%For another view of these results, 
%
Figure~\ref{fig:qfracredshift} shows the quiescent fraction as a function of redshift in bins of stellar mass, separating out the lowest and highest density quartiles. Qualitatively, galaxies have higher quiescent fractions in high-density environments compared to galaxies in low-density environments out to $z\sim2$ and for galaxies with stellar mass, $\log(M/\Msol) \gtrsim 9.8$.  For lower stellar mass galaxies ($8.8 < \log(M/\Msol) < 9.8$), the quiescent fractions are higher in the higher density environment at least to $z \lesssim 1.5$, but at higher redshift, $z \gtrsim 1.5$,  the quiescent fraction shows less dependence on  environment. 

\par The range of quiescent fraction as a function of overdensity and stellar mass is large, ranging from nearly 100\% to less than 1\%.    To illustrate this, we show the quiescent fraction in in log scale instead of linear scale in both  Figure~\ref{fig:qfracmass} and ~\ref{fig:qfracredshift}. This better presents separation of quiescent fractions of galaxies in different environments, and also the relative quiescent fraction in each stellar mass bins.

%
%The increase in the galaxy quiescent fraction in high-density environments might be due to the expectation that more massive galaxies tend to be observed in denser environments, and number of observed all quiescent galaxies increases with %
%stellar mass. We wish to test if the elevated quiescent fraction of galaxies in denser environment is simply due to a relationship between stellar mass and environmental density. 
%
 %So far, we have ignored the fact that higher density regions contain more massive galaxies. 
 %
 Because more massive galaxies tend to exist in higher density environments, it is logical to ask to what extent this drives the trend between mass, redshift, and environment.  This is a reasonable question, as the number density of quiescent galaxies, even at high redshift, increases both with stellar mass and environment (e.g., Papovich et al.\ 2017, in preparation). 
 
 \par To answer this, we perform a similar procedure as \cite{Quadri2012}.  We computed the   mean overdensity of quiescent and star-forming galaxies in narrow 0.25 dex bins in stellar mass (where we expect such narrow bins to have negligible change in overdensity), which show in  Figure~\ref{fig:overdensitymass}. This figure shows evidence that quiescent galaxies have higher overdensity than mass-matched star-forming galaxies (down to the stellar mass limit of each redshift), and this trend exists to $z \sim 2$. These trends extend correlations found by \cite{Quadri2012} to lower stellar masses.  In particular, the density contrast between star-forming and quiescent galaxies is largest for lower mass galaxies, $\log (M/M_\odot) \simeq 9 - 10$. As argued by \cite{Quadri2012}, this suggests that the environment plays a dominant role in galaxy quenching, and here we show this extends to  the lowest stellar masses.  Our analysis supports this assertion,  which we discuss more below.

\subsection{Environmental and Stellar Mass Quenching Efficiencies}
\label{sec:qeff}
To quantify environment and stellar mass in quenching the star-formation activity in galaxies,
we follow the approach of \cite{Peng2010} \citep[which is similar to methods of ][]{vandenBosch2008,Quadri2012,Kovac2014,Lin2014}. We define the environmental quenching efficiency, $\varepsilon_{\mathrm{env}}$, as the fraction of galaxies at a given stellar mass, $M$, that are quenched in excess of those in the lowest-density environment (presumably these are galaxies that would be forming stars in the lowest-density environments, but have had their star-formation truncated due to some physical process related to the environment).   The environmental quenching efficiency is then
\begin{equation}
\label{eq:qeff}
%\varepsilon _{\mathrm{env}} (\delta_{75},\delta_{25}, M)= \frac{f_{q}(\delta_{75}, M) - f_{q} (\delta_{25}, M)}{1 - f_{q}(\delta_{25},M)},
\varepsilon _{\mathrm{env}} (\delta,\delta_{0}, M)= \frac{f_{q}(\delta, M) - f_{q} (\delta_{0}, M)}{1 - f_{q}(\delta_{0},M)},
\end{equation}
where $f_q$ is the quiescent fraction for galaxies with stellar mass $M$ and overdensity, $\delta$.  $\delta_0$ is the overdensity of the low-density reference environment, where we choose $\delta_{0} = \delta_{25}$, i.e., the overdensity demarcating the lowest 25th percentile of the overdensity distribution (see Figure~\ref{fig:overdensity}).   
We note, however, that we are parameterizing non-environmental quenching with stellar mass  because it correlates with other quantities associated with (stellar) mass quenching such as central stellar mass and black hole mass \citep[e.g.,][]{Woo2013,Woo2015,Zolotov2015,Terrazas2016,Tacchella2016a,Tacchella2016b,Woo2017}

\par For galaxies with $\delta \leq \delta_{25}$ our definition of $\varepsilon_{\mathrm{env}}$ explicitly assumes that the environment quenching is negligible (i.e.,$\varepsilon_{\mathrm{env}}(\delta < \delta_0) \approx 0$) for all stellar masses.   This is a reasonable assumption as there is no apparent evolution in the shape of the quiescent galaxy stellar mass function in low-density environments over the redshift and stellar mass range considered here (Papovich et al.\ 2017, in preparation), as would be expected if galaxy quenching correlates only with stellar mass.  For the remainder of this paper we will also denote the (stellar mass dependent) quiescent fraction of galaxies in the lowest and highest-density quartiles as $f_{q}(\delta_{25},M)$ and $f_{q}(\delta_{75},M)$, respectively. 
%
%We will refer to these as the quiescent fractions of galaxies in ``low-density'' and ``high-density'' environments, respectively for convenience. 

\par Similarly, we define the (stellar) mass quenching efficiency $\varepsilon_{\mathrm{mass}}$ as the fraction of
galaxies at a fixed overdensity, $\log(1+\delta)$, that are quenched compared to the star-forming fraction at low masses. Specifically we define the mass quenching efficiency to be, 
\begin{equation}
\label{eq:massqeff2}
\varepsilon _{\mathrm{mass}} (M,M_{0},\delta)= \frac{f_{q}(\delta,M) - f_{q} (\delta,M_0)}{1 - f_{q}(\delta,M_0)},
\end{equation}
where in practice we take the reference mass $M_0$ to be the stellar mass at the completeness limit for a given redshift, and we compute mass quenching efficiency for galaxies with $\delta < \delta_{75}$.

\begin{figure*}
\epsscale{1.11}
%\plotone{twoqeff_lmass}
\plotone{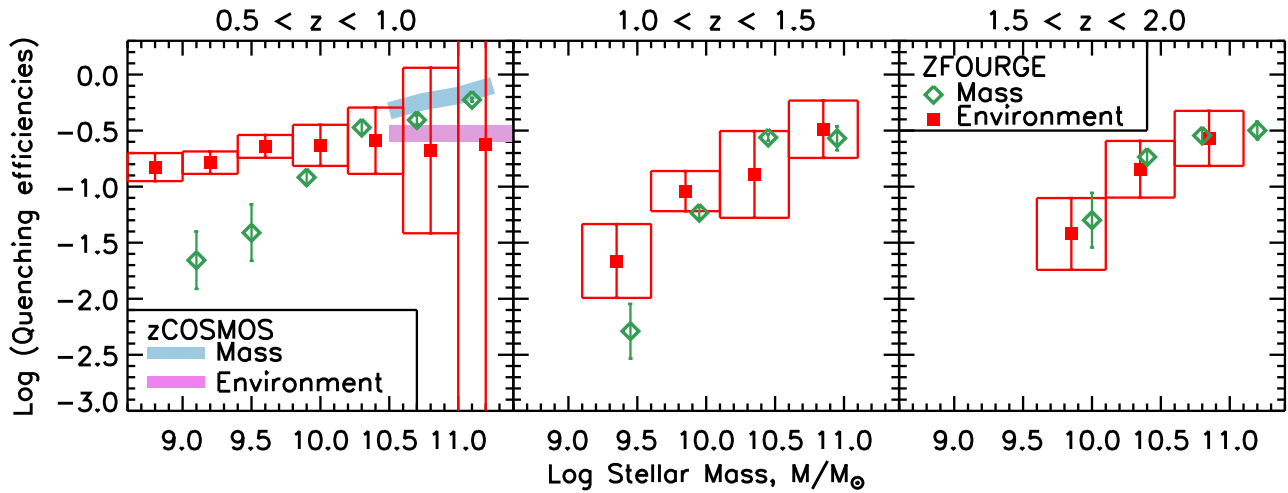}
\epsscale{1.0}
\caption{The environmental quenching efficiency (red squares) and (stellar) mass quenching efficiency (green diamond) versus stellar mass in three different redshift bins. The environmental quenching efficiencies shown here correspond to the highest overdensity quartiles, $\varepsilon_\mathrm{env}(\delta=\delta_{75},\delta_0,M)$.  The box widths show the stellar mass binning, and the box heights (and error bars) indicate $1\sigma$ Poisson uncertainties. The purple and light blue rectangles show the environmental and (stellar) mass quenching efficiencies, respectively, measured in zCOSMOS at $0.3 < z < 0.6$ \citep{Peng2010}. Some error bars are smaller than the size of the data points. The data points are slightly offset for clarity.} 
% see /Users/lalitwadee/ZF_environ/nearest_neighbor/twoqeff_mass_ver3.pro
% see /Users/lalitwadee/ZF_environ/nearest_neighbor/newstat/twoqeff_mass_ver3.pro
\label{fig:qeffmass}
\end{figure*}

\subsection{Dependence of Quenching Efficiencies on Stellar Mass and Redshift}

Figure~\ref{fig:qeffmass} compares the strength of the environmental quenching and (stellar) mass quenching efficiencies as a function of stellar mass for galaxies in the highest density environments ($\delta \ge \delta_{75}$).   At all redshifts, the (stellar) mass quenching efficiency increases with stellar mass.  At $1 < z < 1.5$ and $1.5 < z < 2$, the magnitude of environmental quenching efficiency is on par with the (stellar) mass quenching efficiency:  in the highest density environments roughly half of all galaxies are quenched by the environment.  

\par At lower redshifts, $0.5 < z < 1$, the evolution of environmental quenching efficiency is strongest for lower mass galaxies.  For example, in the mass range, $8.8 < \log (M/M_\odot) < 10$, $\varepsilon_\mathrm{env}$ increases from $<$10\% at $z > 1$ to $\sim$30\% at $z < 1$.    Moreover, at these redshifts the environmental quenching  efficiency dominates over (stellar) mass quenching efficiency for these lower mass galaxies (in the highest density environments).   Therefore, in the highest density environments the majority of quiescent lower-mass galaxies have been quenched by environmental processes  rather than by other processes \citep[see also][]{Hogg2003,vandenBosch2008,Quadri2012}.  Comparing the magnitudes of the environmental and (stellar) mass quenching efficiencies gives an estimate of the effect, which is of order $\varepsilon_{\mathrm{env}}/\varepsilon_{\mathrm{mass}} > 5$ for galaxies with $\log(M/M_\odot) = 8.8-9.8$:  i.e., the environment accounts for the quenching of 5 out of 6 galaxies in this mass range. 

\par Figure~\ref{fig:qeffmass} also shows that at $0.5 < z < 1.0$, the environmental quenching efficiency appears to be nearly independent of stellar mass. At $z > 1$, the environmental quenching efficiency shows a clearer dependence on stellar mass: more massive galaxies experience stronger environmental quenching. This persists at least to $z\simeq 2$ for galaxies with $\log(M/M_\odot) > 9.8$.
%
%at least until the highest galaxy overdensities ($\log(1+\delta^\prime)_3 > 2.2$), where there is evidence that lowest mass galaxies experience lower environmental quenching than more massive galaxies. 
%

\begin{figure*}
\epsscale{1.0}
%\plotone{massqeff_z}
\plotone{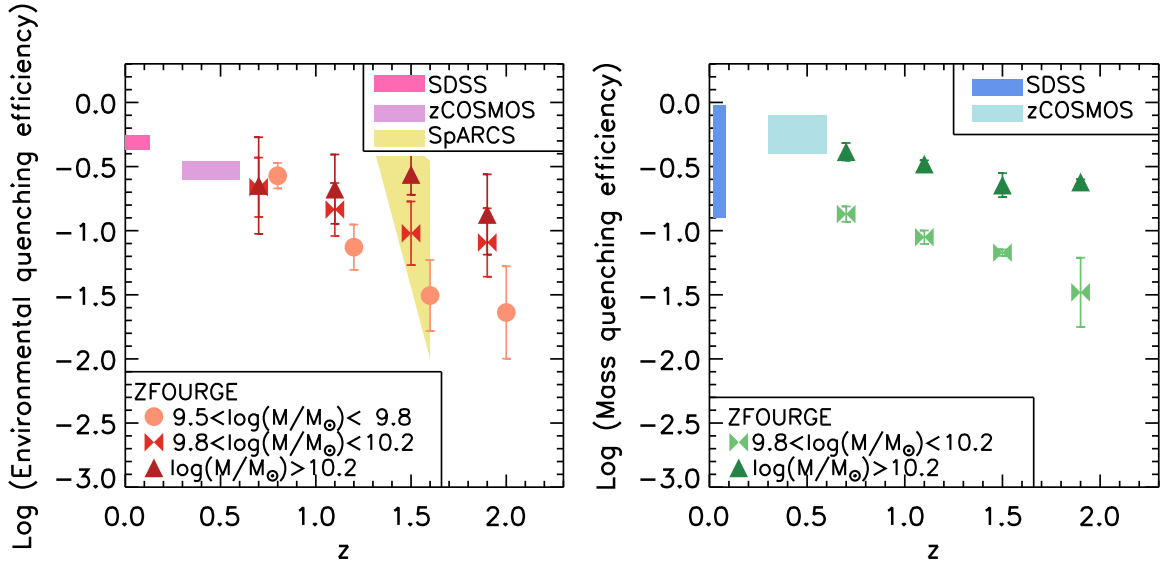}
\caption{The redshift evolution of the environmental quenching efficiency, $\varepsilon_{\mathrm{env}}$ of galaxies in the highest overdensities ($\delta > \delta_{75}$;  Equation~\ref{eq:qeff}; Left panel) and (stellar) mass quenching efficiency ($\varepsilon_\mathrm{mass}$; Equation~\ref{eq:massqeff2}; Right Panel). 
The different symbols denote different bins of stellar mass, as labeled.  The error bars indicate $1\sigma$ uncertainties based on Poisson statistics.  In the left panel, the pink and purple rectangles show the environmental quenching efficiency of galaxies with $9.0 < \log(M/\Msol) < 11.0$ at $0.02 < z < 0.085$ from SDSS and of galaxies with $10.2 < \log(M/\Msol) < 11.0$ at $0.3 < z < 0.6$ from zCOSMOS, respectively \citep{Peng2010}.  The yellow-shaded region shows measurements for galaxy clusters at $0.87 < z < 1.63$ with $\log(M/\Msol) > 10.3$ from the Spitzer Adaptation of the Red-Sequence Cluster Survey (SpARCS) \citep{Nantais2016}. 
In the right panel, the filled rectangles show the (stellar) mass quenching efficiency  of galaxies with $9.0 < \log(M/\Msol) < 11.0$ at $0.02 < z < 0.085$ from SDSS and of galaxies with $10.2 < \log(M/\Msol) < 11.0$ at $0.3 < z < 0.6$ from zCOSMOS, as labeled \citep{Peng2010}.}
% see ~/ZF_environ/nearest_neighbor/massqeff_z_ver2.pro 
% see ~/ZF_environ/nearest_neighbor/twoqeff_z.pro 

% see ~/ZF_environ/nearest_neighbor/newstat/twoqeff_z.pro for the new version

\label{fig:qeffredshift}
\end{figure*} 

\par We explore the evolution with redshift of both the environmental quenching efficiency and the  mass quenching 
efficiency for galaxies in three stellar mass bins: $9.5< \log(M/\Msol) < 9.8$, $9.8< \log(M/\Msol) < 10.2$, and $\log(M/\Msol) > 10.2$ in Figure~\ref{fig:qeffredshift}.
At lower redshifts, $0.5 < z < 1$, the environmental quenching efficiency of galaxies at all masses is $\approx$0.3.  This is generally consistent with that measured with the same (relatively higher) stellar mass at $0.3 < z < 0.6$ from zCOSMOS \citep{Peng2010}.  However, this ``constant'' quenching efficiency is a coincidence of epoch.   At higher redshifts the environmental quenching efficiency of low-mass galaxies decreases and is very low ($\lesssim5\%$) at $z > 1.5$, while for more massive galaxies it remains roughly constant (or possibly slightly declining) out to $z\sim 2$.   The evolution of the environmental quenching efficiency depends both on redshift and stellar mass, and its effects are not separable from stellar mass at higher redshift. 

Comparing to the literature, the environmental quenching efficiency we derive at $1 < z < 1.5$ is modestly lower than that derived for a sample of spectroscopically confirmed galaxy clusters at $0.87 < z < 1.63$ with $\log(M/\Msol) > 10.3$ from the Spitzer Adaptation of the Red-Sequence Cluster Survey (SpARCS) \citep{Nantais2016}. However, this may be expected as the SpARCS sample includes very rich clusters at these redshifts, which include galaxies in even higher overdensities than the  galaxies in our highest density environments in ZFOURGE.  Our result of increasing quiescent fractions in high-density environments implies that the strength of environment quenching efficiency increases with overdensity, it is very reasonable that this efficiency is even higher in the rich environments of galaxy clusters.  

Figure~\ref{fig:qeffredshift} also shows that the strength of (stellar) mass quenching efficiency increases with increasing stellar mass and decreasing redshift. This is consistent with the overall decrease in star-formation activity in galaxies at later cosmic times \citep[e.g.,][]{Madau2014}.

\section{Discussion}
\label{sec:discussion}

\subsection{On the Environmental Impact on Quenching}

Our main result is that there is strong evidence for both (stellar) mass quenching and environmental quenching for galaxies to high redshift.  For massive galaxies, $\log (M/M_\odot) > 10.2$, environmental quenching is evident, and nearly unchanging (or slowly declining), over the redshift range of our sample, $0.5 < z < 2$.    For lower mass galaxies the environmental quenching efficiency evolves strongly with redshift, at least to $\log (M/M_\odot) = 9.5$, where our data are complete.   For such lower-mass galaxies, the environmental quenching declines by roughly an order of magnitude from $z=0.5$ to 2. At our lower redshift range, $0.5 < z < 1.0$, the environmental quenching efficiency dominates over the (stellar) mass quenching efficiency by a factor of $>$5:1 for galaxies with $\log(M/M_\odot) = 8.8-9.8$ (Figure~\ref{fig:qeffmass}). Therefore the majority of low-mass quiescent galaxies are quenched by their environment \citep[][]{Hogg2003,Quadri2012}.
Our result here is consistent with \cite{Geha2012} who found that number of quiescent low-mass galaxies with $7 < \log(M/M_\odot) < 9$ in the field is very low ($<0.06\%$), demonstrating that star-formation of low-mass galaxies  ($\log(M/M_\odot) < 9$) are suppressed by being nearby more massive galaxies \citep[although see][]{Geha2017}.  
In addition, our results are in excellent agreement with recent study by \cite{Guo2017}, who used CANDELS data to measure distance from low-mass galaxies to the nearest massive neighbor galaxies. They found that environmental quenching is dominant quenching mechanism for galaxies with $\log(M/M_\odot) < 9.5$ out to $z\sim1$. At higher redshift, \cite{Guo2017} observed minimal environmental quenching for low-mass galaxies, which is consistent with our finding here, but our observation with ZFOURGE survey provides us sufficient statistics and accurate environment measurement (due to the precise photometric redshifts) to strengthen this result.
%distinguish between minimal environmental quenching and small number statistics.

\par We note that at $0.5 < z < 1.0$, even  in low-density environments, the fraction of  massive quiescent galaxies with stellar mass $\log(M/\Msol) \gtrsim 10.8$  are comparable to those in high-density environment. The observation of quiescent galaxies in voids (low-density environment) has been reported by \cite{Croton2005}. \cite{Croton2008} further compared luminosity function of void galaxies in the 2dF Galaxy Redshift Survey to those from a galaxy formation model built on the Millennium simulation. These authors demonstrated that a population of  quiescent galaxies in low-density environments will arise naturally due to a combination of a shift in the halo mass function in low-density environments and an environment independent star-formation suppression mechanism efficient above a critical halo mass of $M_{\mathrm{vir}}\sim10^{12.5}\msol$ (radio-mode AGN).

\par Some hint of the quenching mechanism comes from the timescales and the evolution in the quenching efficiency.  The lack of significant environmental quenching of low-mass galaxies at $z > 1$ suggests that the quenching timescale is at least 3--5 Gyr (corresponding to the lookback time from $z=1$ to an infall epoch of $z=$ 3 to 6).   This is consistent with quenching times from other studies of environmental processes \citep[e.g.,][]{Peng2010,Tinker2010,Quadri2012,Slater2014,Peng2015,Wetzel2015, Darvish2016,Fossati2016,Guo2017}. Several recent studies \citep[e.g.,][]{Fillingham2015,Peng2015,Davies2016} argue that environmental quenching for galaxies with $\log(M/M_\odot) = 8.0-10.0$ are primarily driven by starvation because quenching timescales and cold gas depletion timescales are comparable.   
For more massive galaxies, $\log(M/M_\odot) \gtrsim 10$, the quenching timescale could be shorter, given that we see higher environmental quenching efficiency for these galaxies even in our highest redshift bin, $1.5 < z < 2.5$.   This suggests a  mass-dependent quenching mechanism, such as ``overconsumption'' \citep{McGee2014}, which arises as more massive (star-forming) galaxies have shorter gas-depletion times (which we discuss more below). 
%
%In addition, the analysis of stellar metallicity in the SDSS galaxies by \cite{Peng2015} show that the strangulation scenario with the quenching timescale of $\sim4$ Gyr is required  to explain the observed stellar metallicity difference between quiescent and star-forming galaxies with $\log(M/M_\odot) < 11$.
Our results support these findings, but with the additional requirement (also discussed below) that the quenching process also transform the morphologies of the quenching galaxies. 

The environmental processes driving the quenching must occur in environments with overdensities comparable to that of our high-density quartile.  The ZFOURGE fields contain some massive groups \citep{Fossati2016}, but no massive, virialized clusters given the cosmological volume contained by the ZFOURGE/CANDELS fields.  Furthermore, our overdensity estimator based on the third nearest neighbor distance measurements are primarily sensitive to group-sized scales \citep{Muldrew2012}.  Therefore the environmental quenching efficiency we measure pertains to physical mechanisms within such environments, and not necessarily to more massive clusters, which may have even stronger environmental quenching efficiency. In addition, even though we do not separate our galaxy sample into central and satellite galaxies in this study, we note that if environmental effects are specific to satellite galaxies, the observed trend here would be even stronger \citep[see][]{Fossati2016}. 

%\vspace{1.0cm}
\subsection{On the Lack of Environmental Impact on Morphology}\label{sec:discussionMorphology}

\par One way to constrain the cause or causes of the environmental quenching is to test if they also affect the morphological structures of the quenched galaxies.  Previous studies have demonstrated a relation between galaxy morphology and star-formation activity, quenched fractions, and implied gas fractions \citep[e.g.,][]{Franx2008,Wuyts2011,Bell2012,Papovich2015}, and quenching is driven by the processes that change morphology and grow black holes \citep[e.g.,][]{Dekel2014,Zolotov2015,Woo2015,Terrazas2016} (non-environmental effects which we refer to as ``(stellar) mass quenching" in this study). 

In contrast, quenching from environmental processes manifests in different ways. One null hypothesis is that environmental quenching has no effect on galaxy morphology. If this is true, then it would suggest that quiescent galaxies in high density environments (which are affected by quenching processes that correlate with stellar mass and environment; Figure~\ref{fig:qeffmass}) would have different morphologies than quiescent galaxies in low density environments (which are affected only by mass quenching); instead, their morphologies would be more similar to the star-forming population in dense environments. We test this hypothesis here by comparing the morphological distributions of quiescent and star-forming galaxies in our different environments.

\begin{figure*}
\epsscale{1.15}
\plotone{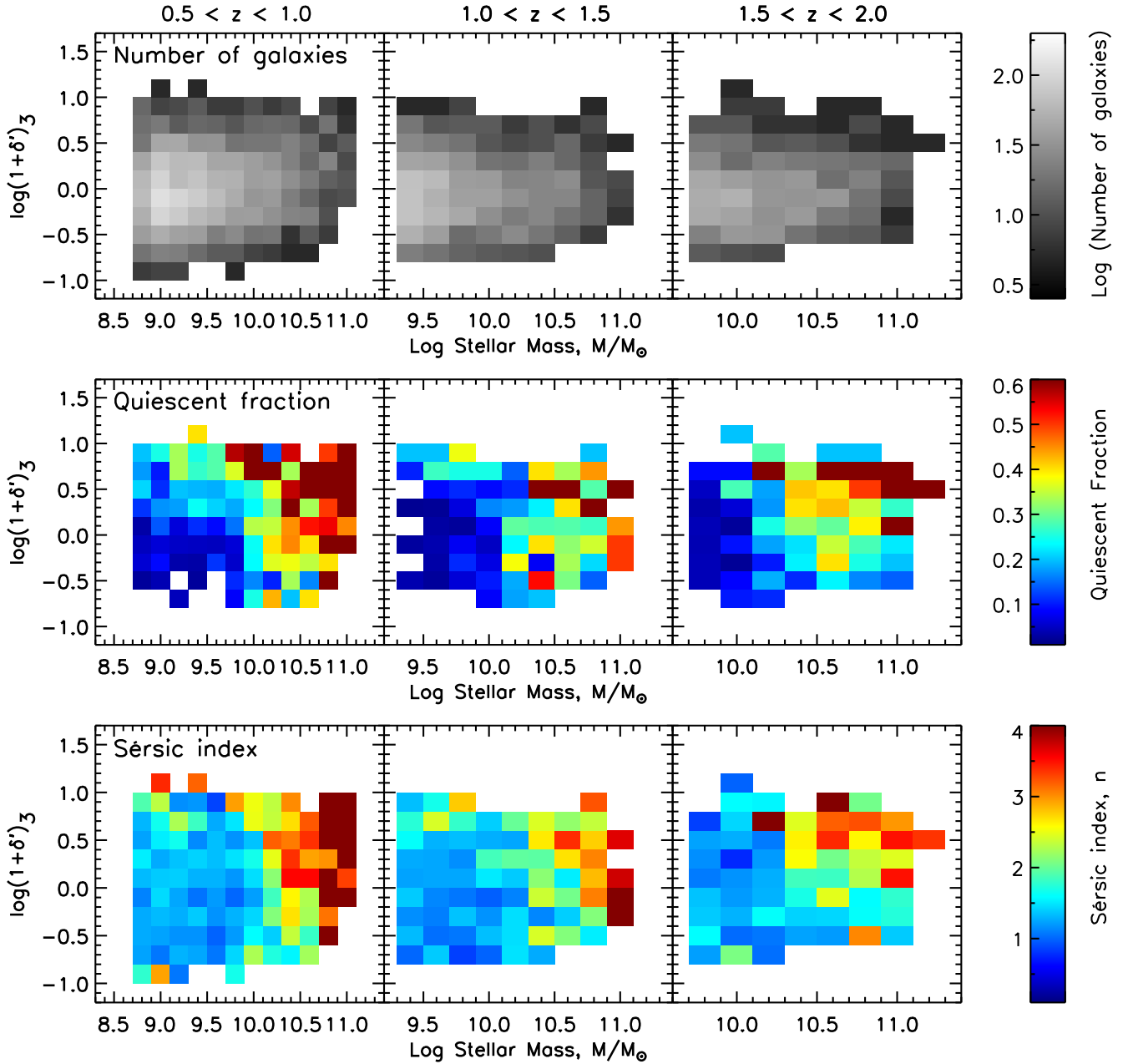}
\caption{\emph{Top:} Number of galaxies as functions of stellar mass and projected local density (environment) in three  redshift bins (from left to right). The number of galaxies is determined in a bin of 0.2 dex in both $\log(M/\Msol)$ and $\log(1+\delta^{\prime})$.  In each panel the color-scaling indicates the number of galaxies (as indicated by the color bar); note that the range of the abscissa changes in each panel to include only galaxies down to the stellar mass completeness in each redshift bin. \emph{Middle:} Quiescent fraction of galaxies as functions of stellar mass and projected local density in three  redshift bins (from left to right). The quiescent fraction of galaxies is determined in a bin of 0.2 dex in both $\log(M/\Msol)$ and $\log(1+\delta^{\prime})$.  In each panel the color-scaling indicates the quiescent fraction (as indicated by the color bar). \emph{Bottom:} The median S\'{e}rsic index of galaxies as functions of stellar mass and projected local density for the same redshift bins as in the top panels. The median S\'{e}rsic index of galaxies is determined in a bin of 0.2 dex in both $\log(M/\Msol)$ and $\log(1+\delta^{\prime})$ as in the top three panels. In each panel the color-scaling indicates the median S\'{e}rsic index (as indicated by the color bar).}
% see /Users/lalitwadee/ZF_environ/nearest_neighbor/onepdelta_mass_fred_sersic_ngal.pro
\label{fig:qfrac2dhist}
\end{figure*} 

\begin{figure*}
\epsscale{1.15}
\plotone{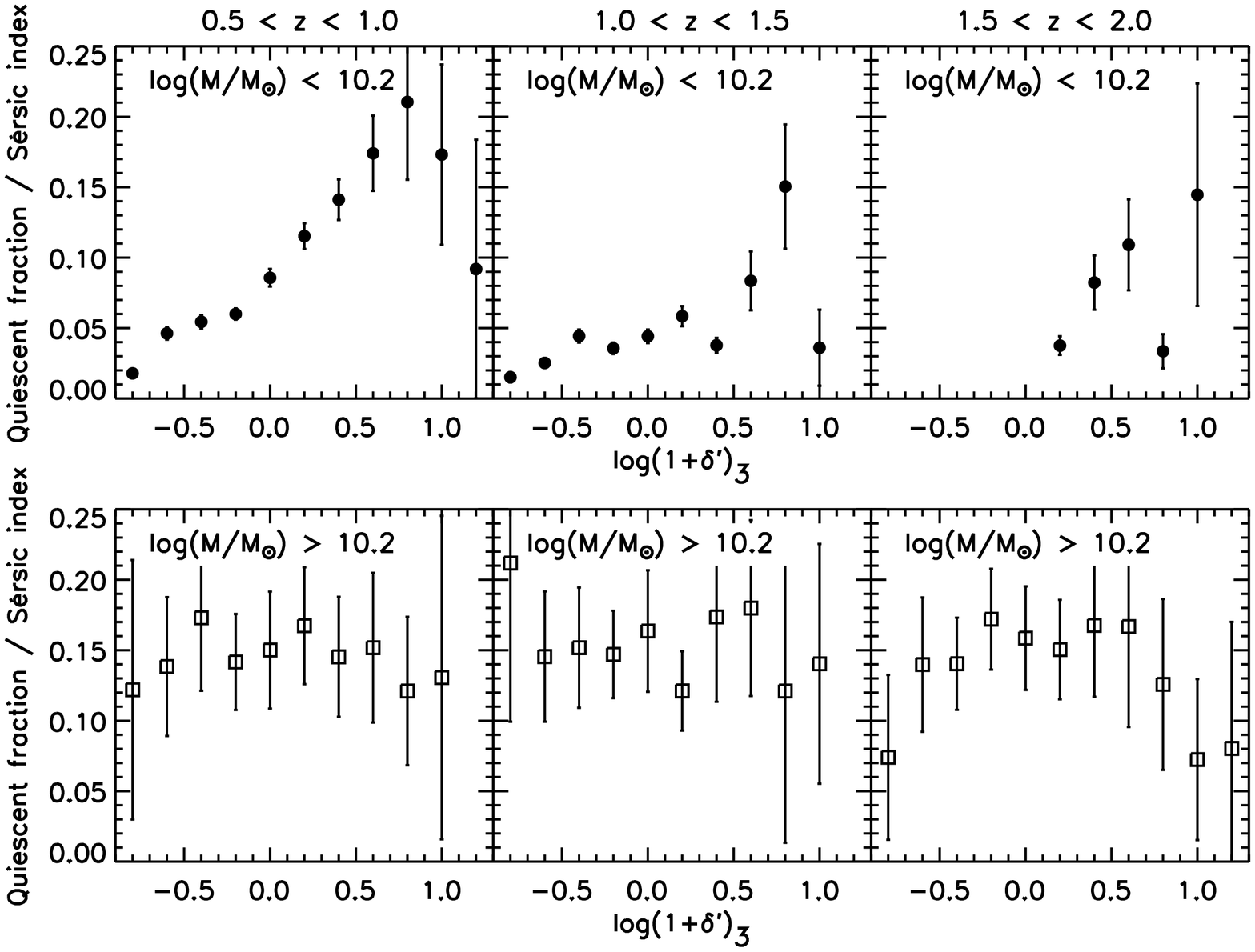}
\caption{\emph{Top:} The ratio of quiescent fraction to the median S\'{e}rsic index of galaxies ($f_q/n$) versus the projected local density (environment) in three redshift bins (from left to right) for  $\log(M/\Msol) < 10.2$ galaxies with stellar mass down to the mass completeness limit in each redshift range. Both quiescent fractions and median S\'{e}rsic index of galaxies are determined in a bin of 0.2 dex in $\log(1+\delta^{\prime})$ as in Figure~\ref{fig:qfrac2dhist}. The error bars indicate $1\sigma$ uncertainties based on Poisson statistics. \emph{Bottom:} Same as the top panels but for the more massive galaxies ($\log(M/\Msol) > 10.2$). }
% see /Users/lalitwadee/ZF_environ/nearest_neighbor/onepdelta_n_qfrac_ver3.pro
\label{fig:qfracsersicratio}
\end{figure*} 

\par \editone{We begin by showing the number of galaxies as functions of stellar mass and projected local density (environment) at $0.5 < z < 1.0$, $1.0 < z < 1.5$, and $1.5 < z < 2.0$ as a function of both $\log(1+\delta^{\prime})_3$ and  $\log(M/\Msol)$ in Figure~\ref{fig:qfrac2dhist} (top panels). We then} show the quiescent fraction of galaxies at the same redshift ranges as a function of both $\log(1+\delta^{\prime})_3$ and  $\log(M/\Msol)$ in Figure~\ref{fig:qfrac2dhist} (middle panels). We observe both environmental quenching and  mass quenching out to $z\sim2.0$ --- the quiescent fraction of galaxies increases with both stellar mass and overdensity as we have shown earlier. 

\par \editone{The middle} panel of Figure~\ref{fig:qfrac2dhist} also shows that for massive galaxies with $\log(M/\Msol) > 10$ the contour of ``constant color'' (quiescent fraction) go from nearly vertical at low redshift to nearly horizontal (the high density regions) at high redshift, demonstrating that we get that 50\% of quenching comes from  mass quenching and 50\% from environmental quenching, even at $z\sim2$ (in the high density regions). This finding is consistent with what we have shown in Figure~\ref{fig:qeffmass}.

\par The bottom panels of Figure~\ref{fig:qfrac2dhist} show the median S\'{e}rsic index of galaxies at $0.5 < z < 1.0$, $1.0 < z < 1.5$, and $1.5 < z < 2.0$ as a function of stellar mass ($\log(M/\Msol)$) and environment ($\log(1+\delta^{\prime})_3$). We find that the distribution of the median S\'{e}rsic index closely resembles that of quiescent fractions of galaxies as shown in the middle panels -- as in previous work (see references above). In this work we see that these quantities closely track each other remarkably well across the all masses, environments and redshifts we probe (this can be seen visually in Figure~\ref{fig:qfrac2dhist}). The similarity of the quiescent fraction and S\'{e}rsic index  distributions is consistent with a picture in which (stellar) mass quenching is reflecting quenching processes that are more directly correlated with morphology, such as bulge-building/compacting mechanisms \citep[e.g.,][]{Lang2014,Zolotov2015,Tacchella2016a,Tacchella2016b,Terrazas2016, Woo2015,Woo2017}.

%in that S\'{e}rsic index and quiescent fraction of galaxies are correlated. 
%quenching of star-formation and morphological change go together, and this correlation persists out to $z=2$, as has been found in other studies (see references above). 

%\par To further quantify how the correlation between quiescent fraction  and S\'{e}rsic index of low mass galaxies changes with environment, 

\par In addition, it is interesting that quiescence and concentrated morphology (S\'{e}rsic index) go together even at high redshift ($z>1.5$) and for the environmental quenching, indicating that only particular galaxies are susceptible to environmental quenching, and those are galaxies which already having high S\'{e}rsic index. These galaxies have time to make it into dense environments, or they collapsed early and have concentrated morphologies. At low redshift it is different, environment affects galaxy star-formation is  independent of their properties. 

%at high redshift the similarity of quiescent fraction and S\'{e}rsic index distributions at high redshift has important implication
 
\par
\begin{figure}
\epsscale{1.2}
\plotone{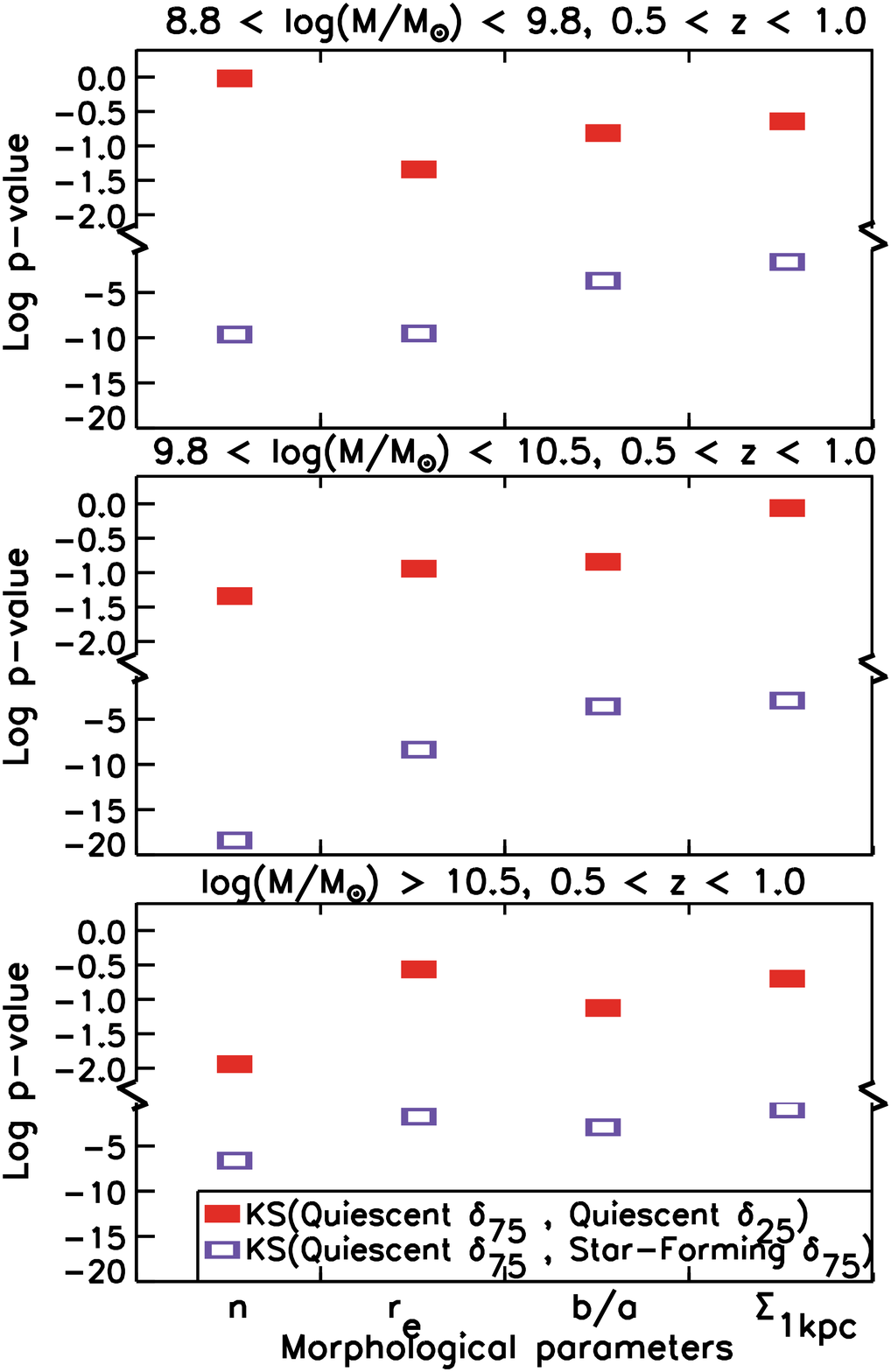}
\caption{The p-values (i.e., likelihoods that the samples have the same parent distribution) from the K-S tests comparing the distributions of four different morphological parameters -- S\'{e}rsic index ($n$), effective radius ($r_e$), axis ratio ($b/a$), and stellar mass surface density in inner 1 kpc ($\Sigma_{1\mathrm{kpc}}$) -- for subsamples of quiescent and star-forming galaxies in different environments.  The top, middle, and bottom panels show different bins of stellar mass for galaxies at $0.5 < z < 1.0$ (as labeled; these are typical -- we do not show higher redshift bins, but they show similar results).  Comparing star-forming and quiescent galaxies in high-density environments (open purple rectangles), in all cases the p-values (\editone{$\lesssim 10^{-3}$}) indicate that we can reject the hypothesis that their morphologies are drawn from the same parent distribution. Performing similar analysis by comparing quiescent galaxies in the highest-density environments to those in the lowest-density environments (filled red rectangles), the p-values indicate that both populations are drawn from the same parent distribution (all p-values are $>$0.05), except for the effective radius distributions of low-mass quiescent galaxies and S\'{e}rsic index distributions of high-mass quiescent galaxies, but \editone{the p-values are only 0.02} (see Section~\ref{sec:morphmethod} and Section~\ref{sec:discussionMorphology} for the discusssion).} 
% see /Users/lalitwadee/ZF_environ/nearest_neighbor/ plotmorphpar_pks_ver3.pro
\label{fig:pksmorphs}
\end{figure}

 There are some indications of deviations from this. As a function of environmental density, for lower mass galaxies with $\log(M/\Msol) < 10.2$ there is some indication that the change in quiescent fraction is faster than the change in galaxy S\'ersic index (at fixed stellar mass), and this exists at all redshifts.  Figure~\ref{fig:qfracsersicratio} shows the ratio of quiescent fraction to the median S\'{e}rsic index of galaxies ($f_q/n$) in 0.2 dex bins of projected local density. At low masses ($\log(M/\Msol) < 10.2$; top panels) the plots show that the ratio of $f_q/n$ is roughly constant for densities, \editone{$\log(1+\delta^{\prime})_3 \lesssim 0$}, at all redshifts, but that the ratio increases at higher overdensity. This is caused by a faster increase in the quiescent fraction while the median S\'{e}rsic index remains roughly constant (or increases slower) as a function of projected local density ($\log(1+\delta^{\prime})_3$) out to $z\sim2$, at least for these low-mass galaxies (this is evident by a close inspection of Figure~\ref{fig:qfrac2dhist}). Our finding here is consistent with \cite{Weinmann2009} who demonstrated that satellite-specific processes mildly enhance concentration of galaxies once they become satellites. This may be taken as some evidence that these low-mass galaxies retain some memory of the morphology of their star-forming progenitors. However, as we discuss below, once galaxies quench (even as a result of their environment), some process also transforms galaxy morphologies on fast time scales as the distributions of the morphological and structural parameters of quiescent galaxies in high and low densities appear highly similar (see Figure~\ref{fig:pksmorphs}).

%Morphologies of quiescent galaxies (quenched by environment) must still transform on fast timescales to produce the results in the K-S tests

\par In contrast, for more massive galaxies ($\log(M/\Msol) > 10.2$; bottom panels of Figure~\ref{fig:qfracsersicratio}) there is no evidence that the ratio of quiescent fraction to the median S\'{e}rsic index of these galaxies increases with environmental density. It is also interesting that there are more quiescent high mass galaxies in dense environments. They already came in with high S\'{e}rsic index, providing tentative evidence for a morphological version of the pre-processing -- the processes that make galaxies concentrated have happened already.
%... this somewhat complicates environment, it's not a super easy reading of field vs. dense, it's rather a continuum of halo masses where the things that are in big halos have for the most part always been in big halos. 

%\par In addition, Figure~\ref{fig:qfracsersicratio} show some tentative evidence that at fixed quiescent fraction of galaxies the low-mass quiescent galaxies in high density environments have lower S\'{e}rsic indices than galaxies in low-density environments.

%%%% I think here is where we can put the Sersic index distribution plot as a funciton of stellar mass and environment that we discussed with Karl G. We will say that that figure shows the distribution of Sersic index as a function of mass and environment.  This plot closely resembles Figure 7 above, in that sersic index and QU fraction are correlated.   

%equivalent to $\gtrsim3\sigma$ significance for Gaussian statistics

%do not reject the hypothesis that they have different morphological distributions 
%In other words, there is no evidence that quiescent galaxies in high-density environments are similar structurally to star-forming galaxies that have been quenched. Nor is there evidence that quiescent galaxies in high density environments are structurally different from those in low density environments.
%
%}

\par We have compared the morphological properties of the galaxy populations using the cumulative distributions of S\'{e}rsic index ($n$), effective radius ($r_e$), axis ratio ($b/a$) \citep[using values from][]{VanderWel2012}, and stellar mass surface density in inner 1 kpc ($\Sigma_{1\mathrm{kpc}}$) for quiescent and star-forming galaxies in the highest and lowest density environments, using the measurements described in Appendix~\ref{sec:morph}. At all redshifts and stellar masses in our sample, the quiescent galaxies have higher S\'{e}rsic indices, smaller effective radii, higher axis ratios, and higher mass surface densities than star-forming galaxies (Figure~\ref{fig:cumuhistallmorphs}), in agreement with previous studies both in the local Universe \citep[e.g.,][]{Bell2008,Fang2013,Omand2014} and at high redshift \citep[e.g.,][]{vanDokkum2011,Wuyts2011,Cheung2012,Bell2012,Szomoru2012,Barro2012,Lang2014,Barro2017}.

Turning to the environmental dependence, we also find that quiescent galaxies in high density environments have different structural parameters than star-forming galaxies in the same environments. This would not be expected if a significant portion of quiescent galaxies in high density environments were recently quenched and retained the morphologies of star-forming galaxies. This is true for all subsamples in stellar mass and redshift that we probe, ($8.5< \log(M/\Msol) < 11.0$ and $0.5 < z < 2.0$). Figure~\ref{fig:pksmorphs} shows a summary of the p-values from K--S tests comparing the distributions of quiescent galaxies in the highest density quartile to those of star-forming galaxies in the highest density quartile at $0.5 < z < 1.0$, and Appendix~\ref{sec:morph} shows the cumulative distributions of the morphological parameters that we tested (S\'ersic indexes, effective radii, axis ratios, stellar mass surface densities) for galaxies at the same redshift. We do not show the higher redshift bins, but we find the same results in all bins of mass and redshift. In all cases the p-values are  $\ll 10^{-2}$. In other words, we can reject the hypothesis that their morphologies are drawn from the same parent distribution.
%with a p-value of $\lesssim10^{-3}$ (equivalent to $\gtrsim 3\sigma$ significance for Gaussian statistics) 

Similarly, there is no evidence (or at best, weak evidence) that quiescent galaxies in high-density environments have different morphologies than quiescent galaxies in low-density environments at any stellar mass or redshift, except at $0.5 < z < 1.0$. At this low redshift we find tentative evidence that low-mass quiescent galaxies ($8.8< \log(M/\Msol) < 9.8$) in high-density environment have larger effective radius than their counterparts in low-density environment, and high-mass quiescent galaxies ($\log(M/\Msol) > 10.5$) have higher S\'{e}rsic index than their counterparts in low-density environment. However, \editone{the p-values of these are only 0.02 (equivalent to $\sim2\sigma$ significance under the assumption of a Gaussian distribution)}. Figure~\ref{fig:pksmorphs} shows a summary of the p-values from K--S tests comparing the distributions of quiescent galaxies in the highest density quartile to those in the lowest density quartile at $0.5 < z < 1.0$. In all cases, we are unable to reject the hypothesis that the morphological distributions are the same  (p-values $>$ $10^{-1.3}$).   

Our results that the morphologies of neither quiescent nor star-forming galaxies depend on environment are generally consistent with previous studies, with some notable exceptions.   Some studies have found differences in the sizes and S\'ersic indexes of galaxies in (lower-density) field and (higher-density) environments at high redshift, but this has mostly been restricted to comparisons between the field and clusters.  \cite{Papovich2012} and \cite{Bassett2013} study the structural and morphological properties of galaxies in a $z=1.62$ proto-cluster and compare those with the field galaxies at the same stellar mass and redshift. Both studies show that the cluster quiescent galaxies have larger average effective sizes compared to field galaxies at fixed mass \citep[see also][]{Cooper2012,Zirm2012,Lani2013,Delaye2014}. In addition, \cite{Bassett2013} found that quiescent cluster galaxies have smaller S\'{e}rsic indices compared to the field galaxies (but this was driven by several quiescent galaxies on the edge of the cluster that may be a rare population of recently quenched star-forming galaxies), whereas the star-forming galaxies in both cluster and field show no difference in their morphologies. 

\par On the other hand, \cite{Newman2014} do not detect a significant difference between the mass-radius relation of the quiescent  galaxies in the cores of clusters (within $R_\mathrm{proj}< 1$~pMpc)  and quiescent field galaxies at $z\sim1.8$. Recently, \cite{Allen2016} studied the dependence of the mass-size relation on environment using field and cluster galaxies at $z\sim1$. The cluster haloes of their sample are on the order of $10^{13}\Msol$, which are comparable to group-sized environments we probe here. \citeauthor{Allen2016} ruled out a size difference of quiescent field galaxies and quiescent cluster galaxies. Similarly, they also showed that the S\'{e}rsic indexes of field quiescent galaxies and cluster quiescent galaxies are consistent. 
Our results are also consistent with \cite{Woo2016}, who compared the specific SFR-$\Sigma_{1\mathrm{kpc}}$ relation for field and satellite galaxies from SDSS with stellar mass $\log(M/\Msol) = 9.75-11.0$, and find that, in a given stellar mass bin, quiescent galaxies have higher $\Sigma_{1\mathrm{kpc}}$ relative to star-forming galaxies by $\sim0.2-0.3$ dex regardless of being field or satellite galaxies.  Therefore, our results add to the growing body of literature that the environment at most weakly affects the morphologies of galaxies when matched in mass, star-formation activity, and redshift.

The lack of evidence for any environmental dependence of the morphological parameters of star-forming or quiescent galaxies has important consequences for the physical effects that drives environmental quenching.  Over the mass and redshift ranges considered here, quiescent galaxies even in high-density environments have very different morphological properties than star-forming galaxies at that epoch, and thus do not appear simply as recently-quenched star-forming galaxies. The implication is that the environmental quenching process transforms galaxy morphologies, and it must do so on timescales comparable to the quenching process.

\par However, it is important to keep in mind the following caveats. We perform the analysis using morphologies as traced by light (``light"-weighted morphology), which may lead to different morphologies as traced by stellar mass \citep[``stellar mass"-weighted morphology][]{Fang2013}. Second, galaxies grow in size with time, but once galaxies is quenched, it stops growing in size at some earlier time. As a result, the quiescent galaxies necessary need to be smaller than star-forming galaxies at a given stellar mass and epoch. Based on this argument, \cite{Lilly2016} demonstrated that a high degree of environmental transformation  might not be needed if one keeps track of morphologies of the progenitor of quiescent galaxies, and links 
mild environmental processes such as stripping of galaxy outer part, disk fading, or removal of dust to reproduce to observed quiescent morphology.
%\subsection{What environmental processes are likely not experienced by galaxies in our study?} 

\subsection{What processes could be driving environmental quenching?} 
\label{sec:processdrivequenching}

%Our results disfavor environmental processes that leave morphological parameters unchanged. This argues against processes that solely affect a galaxy's gas content.
%In addition, the environmental quenching mechanism(s) our observation of the stellar mass dependence of environmental quenching efficiency at high redshift ($z>1$) argues for environment processes which the strength depends on stellar mass, and 

%This argues  against  environment  processes that would only affect a galaxy’s gas content, unless they act in concert with other processes (including disk fading, see~\ref{sec:processdrivequenching} below). Strangulation --- the removal the gas reservoir --- is not expected to affect significantly galaxy morphology \citep[see also][]{vandenBosch2008,Woo2015}. Ram-pressure stripping can remove cold gas from galaxies \citep[e.g.,][]{Vollmer2012,Kenney1999,Oosterloo2005,Chung2007,Sun2007,Abramson2011,Kenney2015}, but again the morphology of a galaxy is not expected to be significantly modified \citep{Weinmann2006,vandenBosch2008}. Moreover, a hot gaseous halo is requirement for ram-pressure stripping to be effective in satellites \citep[e.g.,][]{Larson1980,Balogh2000,Kawata2008,McCarthy2008}, so it is not clear that this can be a dominant mechanism in the lower-mass systems that dominate our study.  

\par The fact that the environmental quenching efficiency evolves with redshift and (at higher redshift) on stellar mass implies that the quenching mechanism itself is correlated with those quantities (namely time and galaxy stellar mass).  This observation is consistent with the  ``overconsumption'' model \citep{McGee2014}, where the environmental quenching time depends on the stellar mass of the satellite. In this model cosmological accretion of gas is halted once a galaxy becomes a satellite of a larger halo, and the decline in star-formation in satellites is then due to the exhaustion of a gas reservoir through star formation and outflows (``starvation''). The timescale on which the galaxy quenches is equal to the total gas available at the point of accretion, divided by the gas consumption rate. 

\par Given the strong correlation between SFR and stellar mass \citep[e.g.,][]{Tomczak2016}, more massive star-forming galaxies have shorter gas-depletion timescales.  In the overconsumption model, \citet{McGee2014} predict that delay times should depend both on galaxy stellar mass and redshift.
 Using this model, \cite{Balogh2016}  showed that the high SFRs of massive galaxies ($\log(M/\Msol) \sim  10.5$) in their sample of groups and clusters at high redshift, $0.8 < z < 1.2$, lead to short delay times at $z\sim1$, consistent with the quenching timescale of at least 3--5 Gyr. This is consistent with the lack of significant environmental quenching of low-mass galaxies at $z > 1$ which we observed here. In addition, \citeauthor{McGee2014} argue that, given the strong redshift evolution of star-formation rate, the quenching timescales should be shorter at $z>1.5$ and is possible even with moderate outflow rates.  

 %In contrast, at $z=0$, the low star-formation and outflow rates imply much longer delay times ($>10$ Gyr) for galaxies at the same stellar mass, in accordance with predictions from the \citeauthor{McGee2014} overconsumption model. 

 \par Nevertheless, the overconsumption model by itself does not account for the differences in the morphological distributions of the galaxies in our study.  Qualitatively,  in its simplest form, the overconsumption model  predicts no morphological evolution of star-formation galaxies (galaxies simply exhaust their gas supply and retain the morphological appearance at infall modulo affects of disk fading, see discussion in \S~5.4 below).   Therefore, while overconsumption can mainly account for environmental quenching of a galaxy at high redshift ($z\gtrsim1$) after it becomes a satellite, alone it probably cannot account for the lack of observed morphological differences in quiescent galaxies in high and low density environments discussed above.

\par At lower redshift, as galaxy specific SFRs declines and associated outflow rates decrease, the quenching time predicted from \citeauthor{McGee2014} overconsumption becomes long ($>10$ Gyr), and other environmental effects that are more closely aligned with dynamical processes in the halo may become more important, and ultimately dominate \citep{Balogh2016}.  This may also drive the environmental quenching efficiency to be more constant with stellar mass at later times ($z \lesssim 0.5-1$), as is observed here and in previous studies \citep[e.g.,][]{Peng2010,Quadri2012,Kovac2014}.

\par At low redshift ($z<1$), there is still clear evidence that the act of becoming quiescent is accompanied by a change in galaxy structure, and this is true even when environmental processes are responsible for quenching star-formation in galaxies. As we discussed above, the environmental quenching processes at low redshift are more likely driven by dynamical processes. Strangulation --- the removal the gas reservoir --- is not expected to affect significantly galaxy morphology \citep[see also][]{vandenBosch2008,Woo2015}. Ram-pressure stripping can remove cold gas from galaxies \citep[e.g.,][]{Vollmer2012,Kenney1999,Oosterloo2005,Chung2007,Sun2007,Abramson2011,Kenney2015},
, but again the morphology of a galaxy is not expected to be significantly modified \citep{Weinmann2006,vandenBosch2008}. Moreover, a hot gaseous halo is requirement for ram-pressure stripping to be effective in satellites \citep[e.g.,][]{Larson1980,Balogh2000,Kawata2008,McCarthy2008}, so it is not clear that this can be a dominant mechanism in the lower-mass systems that dominate our study.  

\par While several studies argue that ram pressure stripping is likely a rapid quenching mechanism, in groups and interacting pairs it is primarily effective in quenching lower mass galaxies ($\log(M/\Msol) < 8.0$) \citep{Slater2014,Davies2015,Weisz2015,Fillingham2016} and is a less relevant quenching mechanism for more moderate mass galaxies \citep[$9.75 < \log M/\Msol < 10$]{Woo2016}.  Given the stellar mass range of galaxies in our samples, their  environmental quenching efficiencies, and expected range of group--halo masses  \citep[e.g.,][]{Fossati2016}, these processes (by themselves) seem unlikely to dominate the overall environmental trends that we observe.

 \subsection{What processes could be driving the environmental morphological transformation?}
 
It may be that multiple environmental processes are at work to quench star-forming galaxies, while others transform their morphologies. 
One candidate for environmental processes that would affect galaxy morphologies are mergers and interactions \citep[similar to ``merging quenching'',][]{Peng2010}, which are expected to be more frequent in denser environments at both $z=0$ and higher redshifts \citep{Fakhouri2009}.  Each merger and interaction can build up the density in the inner kiloparsec of a galaxy \citep{Lake1998}. Such interactions are also shown to be more frequent for higher mass galaxies out to $z=2.5$ \citep{Xu2012,Man2016}, where the interactions could increase gas consumption, redistribute angular momentum, and form spheroids.

\par Frequent galaxy-galaxy encounters also lead to strong tidal torques, which could drive material to galaxy centers, fuel starbursts, build bulges, \citep[e.g.,][]{Sobral2011}, and can also lead to disk stripping.  These could be combined with  disk fading, which enhances the relative importance of the bulge and shifts galaxy morphology toward being more bulge-dominated (higher S\'ersic index and higher $\Sigma_\mathrm{1kpc}$) \citep[e.g.,][]{Carollo2013,Carollo2016}.  To test for disk fading requires morphological measurements weighted by stellar mass (rather than weighted by luminosity, which they are at present). For this, we require spatially resolved optical-near-IR structures in galaxies, which will be possible through forthcoming observations with JWST. A higher rate of (minor) mergers has been invoked to explain the accelerated morphological evolution in galaxies in higher density environments \citep[e.g, at $z\sim 1.6$, ][]{Papovich2012,Rudnick2012,Lotz2013}, and this could explain the  weak evidence that massive quiescent galaxies in the highest density environments have increased S\'ersic indexes compared to massive quiescent galaxies in low density environments  (based on the p-values, Figure~\ref{fig:pksmorphs} and Appendix~\ref{sec:morph}).  While these processes act on galaxies over a range of redshift, they also have the ability to transform galaxy morphologies once they become satellites and would help to explain the structural differences in quiescent and star-forming galaxies in different environments. 
%
%In addition, \cite{Sobral2011} have shown that the median SFRs of H$\alpha$ emitters increase with increasing local surface density (environment), implying that the merger rate is higher.

To summarize, dense galaxy regions are complex environments, and our results suggest that there are multiple processes at work. The redshift and stellar mass evolution of the environmental quenching efficiency favors models where the gas supply is truncated as galaxies become satellites (e.g., starvation), combined with stellar mass dependent star-formation and outflows (e.g., overconsumption). This must be combined with processes such as more frequent interactions and mergers that are prevalent in denser environments and are capable of transforming galaxy morphologies. These processes would naturally connect the quenching timescale with the morphological transformation timescale, which is required to explain the data. This leads to the prediction that massive galaxies in denser environments have more tidal features than those in less dense environments. Again, forthcoming observations with JWST will provide deeper imaging to test our prediction.

\section{Summary}
\label{sec:summary}

We have studied how the local environmental density affects star-formation activity in galaxies using a mass-complete sample to $\log(M/\Msol) > 8.8-9.5$ from deep near-IR ZFOURGE survey at $z = 0.5 - 2.0$.  We measure galaxy overdensities using a Bayesian-motivated estimate of the distance to the 3NN, where the precise photometric redshifts from ZFOURGE ($\sigma_z / (1+z) \lesssim 0.02$) allow us to measure accurately the galaxies in the highest and lowest density environments. We then study the redshift evolution and stellar mass dependence of the quiescent fraction, environmental quenching efficiency, and (stellar) mass quenching efficiency.  The main conclusions of this work are the following: 
\begin{itemize}

\item  The quiescent fraction of galaxies increases in denser environments (greater overdensity). This star-formation-density relation can be traced to at least to $z\sim2.0$, and for galaxies with $\log(M/\Msol) > 9.5$. We show that the star-formation-density relation is not simply the result of a mass-density relation combined with a mass-star-formation relation: even at fixed mass, there is a higher quiescent fraction of galaxies in denser environments, although the significance of this effect is weaker at $z>1.5$. 

\item Both the environmental quenching efficiency and the (stellar)  efficiency evolve with redshift. We observe minimal environmental effects at $z\gtrsim1.5$ ($\lesssim5\%$) for low-mass galaxies ($\log(M/\Msol) < 9.5$), but the strength of environmental quenching increases at later times, eventually dominating over the mass quenching process, particularly at these lower stellar masses. 

\par For more massive galaxies ($\log(M/\Msol) > 9.5$), the environmental quenching efficiency is already significant at high redshift: at $z\sim 2$ it is already $\sim30\%$ in the highest densities and remains roughly constant as a function of redshift to $z\sim 0$. For these massive galaxies, (stellar) mass quenching and environmental quenching are comparable in  high-density environments.

\item The environmental quenching efficiency depends on stellar mass at high redshift, $z > 1$, and \textit{the effects of (stellar) mass quenching and environmental quenching are not separable}.   The environmental quenching mechanisms, particularly for lower-mass galaxies, may be fundamentally different at low and high redshift.  At high redshift, the (stellar) mass-dependence of environmental quenching is qualitatively consistent with a decline in star-formation due to the exhaustion of a gas reservoir through star formation and outflows in the absence of cosmological accretion (overconsumption). On the other hand,  this overconsumption process is less efficient at low redshift, suggesting that external gas stripping process like strangulation may become more important.

\item  The distribution of galaxy morphology as a function of galaxy star-formation activity shows no strong dependence (or at most a weak dependence) on environment.   We find the established relation between star formation and morphology such that quiescent galaxies have higher S\'{e}rsic indices, smaller effective radii, higher axis ratios, and higher mass surface density than mass-matched samples of star-forming galaxies. We do not detect any strong environmental effect on the morphologies of quiescent galaxies (and similarly for star-forming galaxies).  There is the weakest of evidence that at $0.5 < z < 1.0$  in the highest density environments  quiescent massive galaxies have larger S\'ersic indexes and quiescent lower-mass galaxies have larger effective radii than mass-matched quiescent galaxies in low density environments, but the evidence is minimal ($\approx1.5-2\sigma$), and these conclusions are tentative.  

The morphologies suggest that environmental quenching must also transform galaxy morphologies such that there is no observable difference with galaxies in the field.    This is true even for lowest-mass galaxies  ($\log(M/\Msol) = 8.8$) where we expect approximately all such quiescent galaxies to be quenched by their environment.  Therefore, the environmental process responsible for quenching the galaxies also transforms their morphologies such that they no longer share the same parent distribution as mass-matched star-forming galaxies.  

\end{itemize}
\par We argue that the redshift evolution of the mass and environmental quenching favors models that combine ``starvation'' (as galaxies become satellites in larger mass halos) with the exhaustion of a gas reservoir through star-formation and outflows (``overconsumption''). These models must be combined with additional processes such as galaxy interactions, tidal stripping, and disk fading to account for the morphological differences between the quenched and star-forming low-mass galaxy populations

\acknowledgments We wish to thank our collaborators in the ZFOURGE and CANDELS teams for their dedication and assistance, without which this work would not have been possible. We also thank the anonymous referee for a very constructive and helpful report. This work is supported by the National Science Foundation through grants AST 1413317 and 1614668. This work is based on observations taken by the CANDELS Multi-Cycle Treasury Program with the NASA/ESA HST, which is operated by the Association of Universities for Research in Astronomy, Inc., under NASA contract NAS5-26555. This work is supported by HST program number GO-12060. Support for Program number GO-12060 was provided by NASA through a grant from the Space Telescope Science Institute, which is operated by the Association of Universities for Research in Astronomy, Incorporated, under NASA contract NAS5-26555. We acknowledge generous support from the Texas A\&M University and the George P. and Cynthia Woods Institute for Fundamental Physics and Astronomy. LK thanks the LSSTC Data Science Fellowship Program, her time as a Fellow has benefited this work. GGK acknowledges the support of the Australian Research Council through the award of a Future Fellowship (FT140100933). K. Tran acknowledges support by the National Science Foundation under Grant \#1410728.

\begin{appendix}
\section{Measuring Galaxy Overdensities with Photometric Redshifts}
\label{sec:densitysimulation}

In this work we use a variation of the distance to the 3NN as a measure of the galaxy environment. We derive the 3NN from the ZFOURGE photometric redshifts and here we quantify how well the overdensity derived from the 3NN reproduces the true overdensity as measured in spectroscopic redshift surveys.  

\par We use a mock galaxy catalog based on the semi-analytic model from \cite{Henriques2015}, which is the Munich galaxy formation model updated to the Plank first-year cosmology. \citeauthor{Henriques2015} modify the treatment of baryonic processes to address  the overabundance of low-mass galaxies and quiescent galaxies. For our purposes the details of galaxy formation and feedback in the mock are less important than the actual redshifts (which include both the cosmological expansion and peculiar velocity).  We take all galaxies in the mock down to the stellar mass limit of our ZFOURGE survey at every redshift.   We then perturb the redshift from the mock catalog by a random number selected from a normal distribution with $\sigma_z$ equal to the uncertainty of the ZFOURGE photometric redshift \citep[$\sigma_z = 0.01(1+z)$ to $0.02 (1+z)$, depending on the galaxy mass and magnitude, see][]{Straatman2016}.   

We then calculate the distance to the $N$th nearest neighbor, with $N$=1, 2, 3, \ldots, 7 in two ways.  First, we use the ``true'' redshifts from the mock (which include the peculiar velocity) to measure the distance to the nearest neighbor within a cylindrical volume of length (in the radial dimension) corresponding to $\Delta v = \pm 2100 \unit{km~s}^{-1}$, which serves as an estimate of the density measured in a spectroscopic survey.   We then repeat the  measurement, but using the perturbed redshifts as an estimate of the density measured by a ZFOURGE-like photometric redshift survey.  

We then compute two estimates of the local surface density of galaxies derived from the $N$th nearest neighbor, first with the standard method, $\Sigma_N = N (\pi d^2_N)^{-1}$, where $d_N$ is the distance to the $N$th nearest neighbor.   Second, we use the Bayesian-motivated estimate of the local surface density derived to the $N$th nearest neighbor, $\Sigma_N^\prime$, which uses the density information of the distances to all neighbors $\le N$ \citep{Ivezic2005,Cowan2008} defined in Equation~\ref{eq:bayesian} above. From both estimates of the local surface density, we compute the overdensity, $(1+\delta)_N$, using Equation~\ref{eq:overdensity} above. 

\begin{figure}
\epsscale{1.0}
\plotone{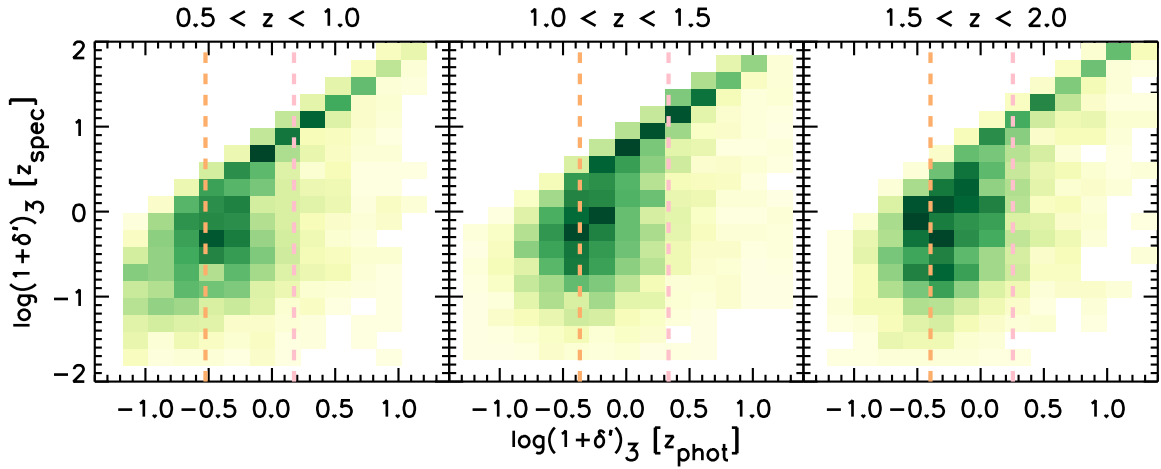}
\caption{Simulated density measurements based on photometric redshifts ($\log(1+\delta^{\prime})_{N=3}~[z_\mathrm{phot}]$; $\Delta z = 0.02(1+z)$) versus measurements based on spectroscopic redshifts ($\log(1+\delta^{\prime})_{N=3}~[z_\mathrm{spec}]$; $\Delta v =2100$ km/s) for the Bayesian-motivated $N$=3 (3rd nearest neighbor [3NN]) in three redshift ranges. The orange and pink vertical dashed lines indicate the lowest and top quartiles (the 25th and 75th percentiles of $\log(1+\delta^{\prime})_{N=3} [z_\mathrm{phot}]$ distribution), which we used to specify low- and high-density environments. There is a strong correlation between both density measurements -- at all redshifts galaxies identified in the highest (lowest) density quartiles using photometric redshifts with this precision recover those galaxies in the highest (lowest) density quartiles as defined using spectroscopic redshifts. }
% see /Users/lalitwadee/ZF_environ/nearest_neighbor/Henriques2015a/calcsigma_1pdelta_zpturbed_ver2
\label{fig:onepdeltamock}
\end{figure} 
%The orange and pink vertical solid lines indicate the lowest and top quartiles of $\log(1+\delta^{\prime})_{N=3} [z_\mathrm{spec}]$ distribution.

\begin{figure}
\epsscale{1.0}
\plotone{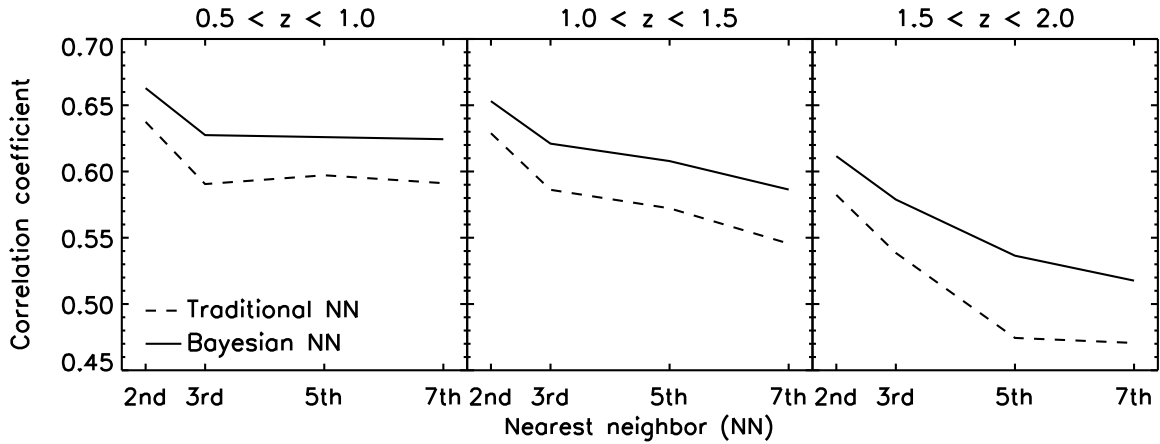}
\caption{Spearman's correlation coefficient between photometric environment measurements and spectroscopic environment measurements as a function of the $N$th nearest neighbors with $N=2,3,5,7$ for mock galaxy sample at $0.5 < z < 1.0$ (left), $1.0 < z < 1.5$ (middle), and $1.5 < z < 2.0$ (right). The Bayesian-motivated and traditional $N$th nearest neighbors are shown as solid and dashed lines, respectively.}
% see /Users/lalitwadee/ZF_environ/nearest_neighbor/Henriques2015a/calcsigma_1pdelta_zpturbed_ver2.pro 
\label{fig:spearman}
\end{figure} 

For convenience, we define $\log(1+\delta^\prime)_N~[z_\mathrm{phot}]$ as the density measured for galaxies in the mock using the ZFOURGE-like photometric redshift uncertainties.  We also define $\log(1+\delta^\prime)_N~[z_\mathrm{spec}]$ as the density measured in the mock using ``true'' redshifts (which include the cosmological and peculiar redshift in the mock).  
Figure~\ref{fig:onepdeltamock} shows the overdensities for the Bayesian 3NN for the mock catalog, $\log (1+\delta^\prime)_{N=3}[z_\mathrm{spec}]$, compared to those derived from the photometric redshifts, $\log(1+\delta^\prime)_{N=3}[z_\mathrm{phot}]$. The relations are clearly correlated, with galaxies at low and high overdensity in the spectroscopic survey generally displaying low and high overdensity when measured in the ZFOURGE--like survey.  However, there are clear examples of mismatch, for example there is a tail of objects with low overdensity, as defined in the spectroscopic-quality data, that have measured high overdensity in the ZFOURGE--like data.  This tail is caused by redshift errors and chance projections of unassociated objects along the line of sight, and is consistent with the findings of \citet{Cooper2005} that photometric redshifts can wash out structure.   

%\todo{I can imagine a referee liking the tests described here, but they could (rightly) say that we don't know even for the spectroscopic-like mock catalog if objects with measured high overdensity are in high-density environments.  After we get everything else done in the paper, we should start making plots of overdensity, $\log(1+\delta)[dv]$ and $\log(1+\delta)[dz]$ against halo mass of the *central* (so if a galaxy is a central, use its halo mass, and if a galaxy is a satellite, use the central's halo mass).   I think we have that information in the Henriques SAM and it would help show that the overdensity scales with halo mass (or something).  Let's plan to do this (next, after we're circulating the paper to all the coauthors!)} 

\par To quantify the accuracy of the overdensity as measured by the ZFOURGE-like dataset, we first measured the Spearman's correlation coefficient for the overdensities measured in the spectroscopic-like and ZFOURGE-like datasets for the overdenstiy calculated using the $N=2$, 3, 5, 7th nearest neighbor, for the standard and Bayesian methods described above.   Figure~\ref{fig:spearman} shows the correlation coefficients as a function of $N$ for the mock galaxies.  We find that for all redshift bins from $z=0.5$ to 2.0, the Bayesian density estimators have higher Spearman's correlation coefficient relative to the traditional nearest neighbors regardless of number of nearest neighbors, indicating that the Bayesian density estimator is better correlated with the spectroscopic density estimator relative to the standard $N$th nearest neighbors.  The two-sided significance (p--value) of the Spearman's correlation coefficient for all three redshift bins are \editone{zero}, implying very strong correlation.   Second, we find that the correlation increases with lower $N$th nearest neighbors.  

\par Second, we computed the completeness and contamination in low- and high-density environments derived using the ZFOURGE-like mock survey. Our goal is to derive robust samples of galaxies in low-density (high-density) environments that are relatively ``pure'' in that they have a low contamination fraction of galaxies in high-density (low-density) environments misclassified by our method.    Specifically, we will define samples of galaxies in high density and low density environments based on the ranked quartiles \editone{computed using a spline linear regression implemented with the cobs package as we did for our analysis of the ZFOURGE galaxy sample} (where we define galaxies in the top density quartile as ``high density'' and those in the bottom density quartile  as ``low density''.)   

Selecting galaxies in the top (highest) density quartile in $\log(1+\delta^\prime)_N~[z_\mathrm{phot}]$, using the 3NN ($N$=3) we recover $>$80\% of galaxies that are also in the top density quartile in $\log(1+\delta^\prime)_N~[z_\mathrm{spec}]$ (see Figure~\ref{fig:onepdeltamock}; the completeness declines for higher values of $N$).   Of the galaxies in the top density quartile in $\log(1+\delta^\prime)_N~[z_\mathrm{spec}]$ that we miss, more than half are in the next (3rd) quartile in $\log(1+\delta^\prime)_N~[z_\mathrm{phot}]$.   More importantly, the contamination is low.    The fraction of galaxies in our top density quartile in $\log(1+\delta^\prime)_N~[z_\mathrm{phot}]$ that are actually in the lowest or 2nd-lowest  density quartiles in $\log(1+\delta^\prime)_N~[z_\mathrm{spec}]$ is $\sim$15\% (using $N$=3, i.e., the 3NN;  the contamination increases for larger choices of $N$ and also increases when using the non-Bayesian estimator for the nearest-neighbor distance). This is acceptable as our goal is to identify a relatively pure sample of galaxies in high density environments, which we achieve.  In other words, we lose about one-third of galaxies that should be in our top density quartile, but the majority ($>$85\%) of galaxies in our top density quartile are truly in high density environments  as measured by $\log(1+\delta^\prime)_N~[z_\mathrm{spec}]$) (i.e., there is a low incidence of chance alignments of galaxies in projection compared to real, physically associated galaxies).   

Selecting galaxies in the bottom (lowest) density quartile in $\log(1+\delta^\prime)_N~[z_\mathrm{phot}]$, we recover  $\sim$80\% of galaxies that are also in the lowest density quartile in $\log(1+\delta^\prime)_N~[z_\mathrm{spec}]$ for $N$=3 (see Figure~\ref{fig:onepdeltamock}).  (As for galaxies in high density environments, we find the completeness decreases for higher $N$ and when using the non-Bayesian estimator for the nearest neighbor distance).   The sample of galaxies in the lowest density quartile measured by $\log(1+\delta^\prime)_N~[z_\mathrm{phot}]$ is very pure in that it contains almost no contamination of galaxies in high density environments in $\log(1+\delta^\prime)_N~[z_\mathrm{spec}]$ (see Figure~\ref{fig:onepdeltamock}):   we find the contamination of sources that are in (from the spectroscopic redshift survey) the highest density or 2nd highest density quartile is $15$\% (the contamination again increases for larger values of $N$ and when using the non-Bayesian estimator for the nearest neighbor distance).    Again, this is acceptable as it says that the majority ($80$\%) of galaxies identified in our lowest density quartile in $\log(1+\delta^\prime)_N~[z_\mathrm{phot}]$ are in low density environments. 

Taking the information about the correlation coefficients, completeness, and contamination together, this provides justification that the overdensity derived from the Bayesian 3NN density estimator accurately recovers galaxies in high and low densities with a strong correlation between measured density and true density.  We also choose the $N$=3rd nearest neighbor as it allows higher completeness and lower contamination (see above).  A further advantage of using the 3NN  as a density indicator is that it has also been shown to provide a faithful measure of the local environmental on scales of galaxy and galaxy group halos, which is appropriate for our study here \citep{Muldrew2012}.  Therefore, we adopt the Bayesian 3NN as our density measure for the study here.

\section{Stellar mass surface density in inner 1kpc and cumulative distributions of structural Morphological Parameters}
\label{sec:morph}

\begin{figure*}
\epsscale{0.9}
\plotone{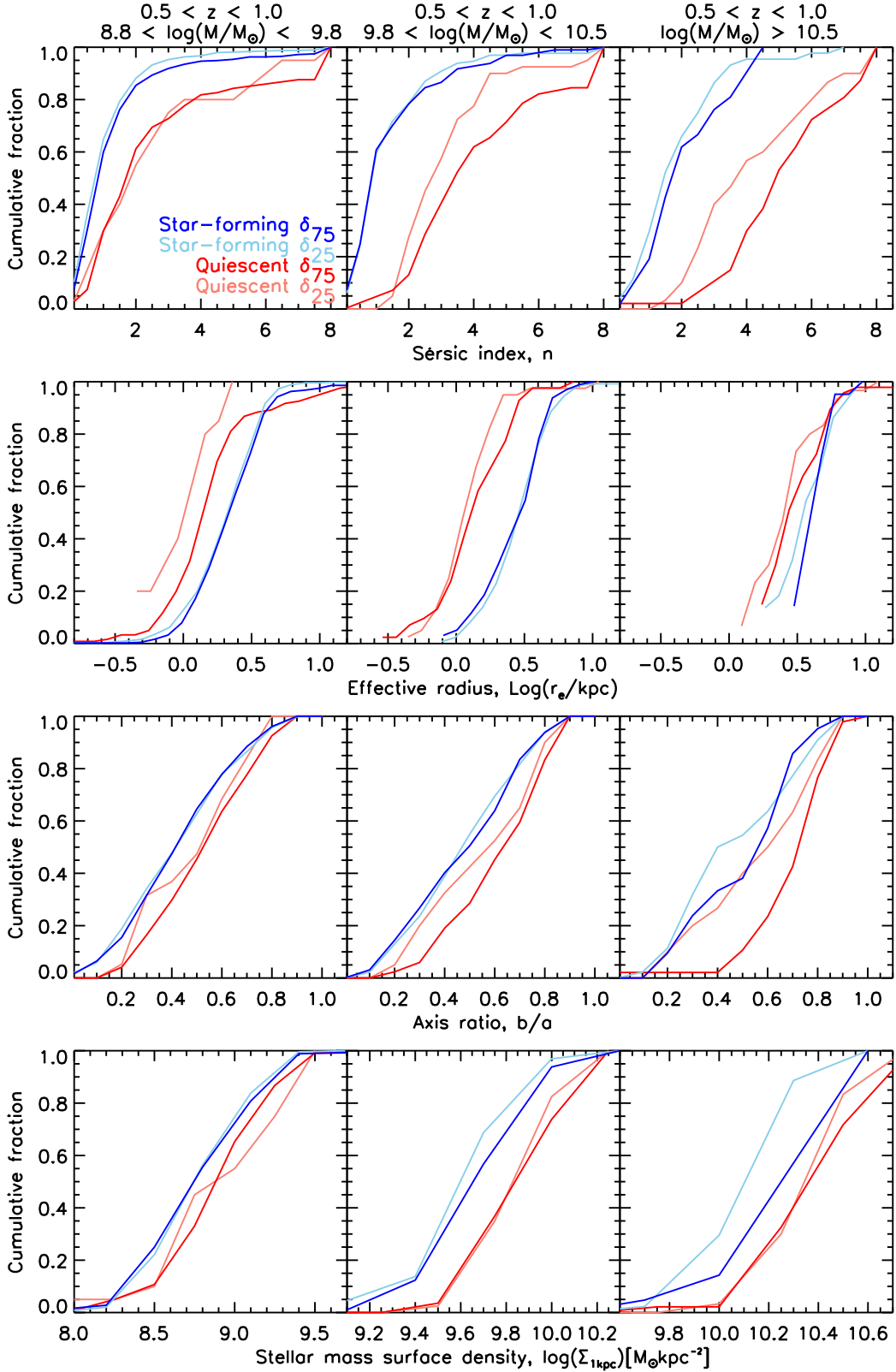}
\caption{The cumulative distribution of S\'{e}rsic index (top row), effective radius (second row), axis ratio (third row), and stellar mass surface density in inner 1 kpc (bottom row) for quiescent galaxies in the lowest-density quartile ($\delta_{25}$; light-red), quiescent galaxies in the highest-density quartile ($\delta_{75}$; red), star-forming galaxies at $0.5 < z < 1.0$ in the lowest-density quartile ($\delta_{25}$; light blue), and  star-forming galaxies in the highest-density quartile ($\delta_{75}$;blue) in three stellar mass ranges. 
There is no significant evidence for a difference
in distributions of any of the four morphological parameters between quiescent (and star-forming) galaxies in low- and high-density environment, except for the effective radius distributions of low-mass quiescent galaxies and S\'{e}rsic index distributions of high-mass quiescent galaxies but only at $2\sigma$ level significance (see Section~\ref{sec:discussionMorphology} for the discusssion).}
% see /Users/lalitwadee/ZF_environ/nearest_neighbor/ plotcumuhist_all.pro
\label{fig:cumuhistallmorphs}
\end{figure*} 

%In our analysis, we study the morphological differences between quiescent and star-forming galaxies in different environments and as a function of stellar mass.   The majority of galaxies in our sample fall within the CANDELS coverage from HST/WFC3, with effective semi-major axis, $a_\mathrm{eff}$,  and S\'{e}rsic index, $n$, measured by \citet{VanderWel2012} using the HST/WFC3 F160W ($H_{160}$)--band imaging.   We cross--matched the sources in our catalog with those of \citeauthor{VanderWel2012}.  We further define the circularized effective radius as $r_\mathrm{eff} = a_\mathrm{eff} \sqrt{q}$, where $a_\mathrm{eff}$ is the effective semi-major axis and $q = b/a$ is the ratio of the semi-minor to semi-major axis. 
In addition to comparing the three morphological parameters,  S\'{e}rsic index, effective semi-major axis, axis ratio (described in Section~\ref{sec:morphmethod}) of quiescent and star-forming galaxies in low- and high-density environments. We also consider correlations with the stellar mass surface density within the inner 1 kpc, $\Sigma_\mathrm{1kpc}$. We calculate $\Sigma_\mathrm{1kpc}$ following the procedure described by \cite{Bezanson2009} and \cite{Whitaker2016} using the galaxies' best-fit S\'{e}rsic indexes ($n$) and circularized effective radii ($r_\mathrm{eff}$). In brief, we assume isotropic spherical galaxies with surface luminosity profiles following the S\'{e}rsic profile to perform an Abel Transform to deproject the circularized, three-dimensional light profile:
\begin{equation}
\rho\left(\frac{r}{r_e}\right) = \frac{b_n}{\pi}\frac{I_o}{r_e}\left(\frac{r}{r_e} \right )^{1/n-1}\int_{1}^{\infty}\frac{\exp[-b_n(r/r_e)^{1/n}t]}{\sqrt{t^{2n}-1}}dt
\end{equation}
We convert the total luminosity to total stellar mass, assuming that mass follows the light, and there are no strong color gradients. We follow \cite{vanDokkum2014} by applying a small correction to these stellar masses to take into account the different between the total magnitude in the photometric catalog and the total magnitude implied by S\'{e}rsic fit \citep[see][]{Taylor2010}. Finally, we calculate the stellar mass surface density in inner 1 kpc by numerically integrating the following equation
\begin{equation}
\Sigma_{1\mathrm{kpc}} = \frac{\int_{0}^{1~\mathrm{kpc}} \rho(r)r^2 dr }{  \int_{0}^{\infty}\rho(r)r^2 dr}\frac{L_{\mathrm{model}}}{L_{\mathrm{phot}}}\frac{M_{\mathrm{phot}}}{\frac{4}{3}(1~\mathrm{kpc})^3}
\end{equation}
where $M_{\mathrm{phot}}$ is the stellar mass of the galaxy from the ZFOURGE catalogs, $L_{\mathrm{phot}}$ is the total, aperture-corrected luminosity from the ZFOURGE catalogs in the bandpass corresponding to the S\'{e}rsic profile measurement   ($H_{160}$), and $L_{\mathrm{model}}$ is the total luminosity as measured from integrating the best-fit S\'{e}rsic profile.

\par Figure~\ref{fig:cumuhistallmorphs} shows the cumulative distributions of  S\'{e}rsic index ($n$), effective radius ($r_e$), axis ratio ($b/a$), and stellar mass surface density in inner 1 kpc ($\Sigma_{1\mathrm{kpc}}$) for quiescent galaxies and star-forming galaxies at $0.5 < z < 1.0$ in the lowest and highest overdensity quartiles for three stellar mass bins.  In all cases, we find that there is no statistical difference in the distributions of quiescent galaxies in high density environments and quiescent galaxies in low density environments.  Similarly, we find that the morphology distributions of star-forming galaxies and quiescent galaxies in the highest density regions are dissimilar (therefore, quiescent galaxies in high density regions do not have the morphologies of (recently quenched) star-forming galaxies.  We use these distributions in \S~\ref{sec:discussionMorphology} (see also Figure~\ref{fig:pksmorphs}). Note that we do not show our  higher redshift bins in the figures, but  we find the same results in all bins of mass and redshift.  
\end{appendix}
%\bibliography{manuscript2_aas61}
\bibliography{references}

\end{document}